 \documentclass[final,5p,times,twocolumn,authoryear]{elsarticle}

\usepackage{amssymb}
\usepackage{lipsum}
\usepackage{url}

\usepackage{hyperref}
\hypersetup{
    colorlinks=true,
    linkcolor=blue,
    filecolor=magenta,      
    urlcolor=cyan,
    citecolor=blue,
}
\newcommand{\js}{} % change to \textit to put abbrevs. in italics

\newcommand\aj{\js{AJ}}%        % Astronomical Journal 
\newcommand\araa{\js{ARA\&A}}%  % Annual Review of Astron and Astrophys 
\newcommand\apj{\js{ApJ}}%      % Astrophysical Journal 
\newcommand\apjl{\js{ApJL}}     % Astrophysical Journal, Letters 
\newcommand\apjs{\js{ApJS}}%    % Astrophysical Journal, Supplement 
\newcommand\aap{\js{A\&A}}%     % Astronomy and Astrophysics 
\newcommand\mnras{\js{MNRAS}}%   % Monthly Notices of the RAS 
\newcommand\pasp{\js{PASP}}%     % Publications of the ASP 
\newcommand\actaa{\js{AcA}}%  % Acta Astronomica
\newcommand{\red}[1]{\textcolor{red}{#1}}

\journal{New Astronomy}

\begin{document}

\begin{frontmatter}



\title{{\it Gaia's} binary star renaissance}


\author[first]{Kareem El-Badry}
\ead{kelbadry@caltech.edu}
\affiliation[first]{organization={California Institute of Technology},%Department and Organization
            addressline={1216 E California Blvd}, 
            city={Pasadena},
            postcode={91125}, 
            state={CA},
            country={USA}}

\begin{abstract}
Stellar multiplicity is among the oldest and richest problems in astrophysics. Binary stars are a cornerstone of stellar mass and radius measurements that underpin modern stellar evolutionary models. Binaries are the progenitors of many of the most interesting and exotic astrophysical phenomena, ranging from type Ia supernovae to gamma ray bursts, hypervelocity stars, and most detectable stellar black holes. They are also ubiquitous, accounting for about half of all stars in the Universe. In the era of gravitational waves, wide-field surveys, and open-source stellar models, binaries are coming back stronger than a nineties trend. Much of the progress in the last decade has been enabled by the {\it Gaia} mission, which provides high-precision astrometry for more than a billion stars in the Milky Way. The {\it Gaia} data probe a wider range of binary separations and mass ratios than most previous surveys, enabling both an improved binary population census and discovery of rare objects. I summarize recent results in the study of binary stars brought about by {\it Gaia}, focusing in particular on developments related to wide ($a \gtrsim 100$\,au) binaries, evidence of binarity from astrometric noise and proper motion anomaly, astrometric and radial velocity orbits from {\it Gaia} DR3, and binaries containing non-accreting compact objects. Limitations of the {\it Gaia} data, the importance of ground-based follow-up, and anticipated improvements with {\it Gaia} DR4 are also discussed.
\end{abstract}



\begin{keyword}
%% keywords here, in the form: keyword \sep keyword, up to a maximum of 6 keywords
binaries: visual \sep binaries: spectroscopic \sep binaries: astrometric \sep stars: black holes \sep white dwarfs



\end{keyword}

\end{frontmatter}

%\tableofcontents



\section{Introduction}
\label{introduction}

Binary stars have long played a foundational role in astrophysics. They underpin precision measurements of stellar physical parameters, enable robust tests of general relativity, and give rise to an extraordinary zoo of observational phenomenology. Millennia after their discovery \citep[e.g.][]{Jetsu2015}, binaries remain at the heart of many of the interesting open questions in astrophysics: binary evolution modeling is key for understanding the origin of gravitational wave events, the spectral energy distributions of high redshift galaxies, and the demographics of exoplanets in the solar neighborhood.

Astrometry has played a particularly important role for binary star astronomy. Painstaking measurements of the relative positions of two stars in resolved optical pairs over the course of decades allowed \citet{Herschel1803} to infer that most of the pairs he studied were orbiting one another. By monitoring the motion of Sirius and Procyon over a century, \citet{Bessel1844} realized that, despite having no visible companions, these stars did not move in straight lines. He correctly deduced that the stars' ``special motions'' were the result of the gravity of massive unseen companions, now known to be white dwarfs.

The {\it Gaia} mission \citep{Prusti2016, Brown2018, GaiaCollaboration2023} brought astrometry to the masses. {\it Gaia} has revolutionized the study of binary stars -- and indeed, of all stars -- by delivering precise and accurate distance and proper motion measurements for hundreds of millions of sources in the Milky Way. {\it Gaia's} promise for binary population demographics and the discovery of rare objects was recognized well before launch  \citep[e.g.][]{Perryman2001, Soderhjelm2004, Fuchs2005}. The mission builds on the legacy of {\it Hipparcos} \citep{Perryman1997, ESA1997} -- which allowed for discovery and characterization of binaries in the solar neighborhood \citep{Soderhjelm1999} -- but {\it Gaia} looks $100$ times deeper into the Galaxy than {\it Hipparcos} did, opening up a discovery space $10^6$ times larger.

This review summarizes major developments in binary star astrophysics that have been enabled by {\it Gaia}, and the state of the field as of early 2024. I will neither attempt to cover the full history of the field (which is long, and mostly predates {\it Gaia}) nor provide a comprehensive list of results. %: binary star studies have a long history, mostly predating {\it Gaia}, and the explosion of results in the last decade would make a comprehensive summary tedious. 
Instead, I will attempt to highlight the most significant new findings and open questions, some of which will likely be answered by the {\it Gaia} data in the coming years, and to provide references for interested readers to dig deeper.  For a broader perspective, I refer to previous reviews by \citet{Abt1983}, \citet{Duchene2013}, \citet{Moe2017}, \citet{Offner2023}, and \citet{Chen2024}.

None of the work discussed here would be possible without an enormous investment of time and resources by the {\it Gaia} data processing and analysis consortium (DPAC) to provide a well-documented, high-quality data product. The community owes DPAC!

\section{A many-pronged approach to the binary census}

\begin{figure*}[!ht]
    \centering
    \includegraphics[width=\textwidth]{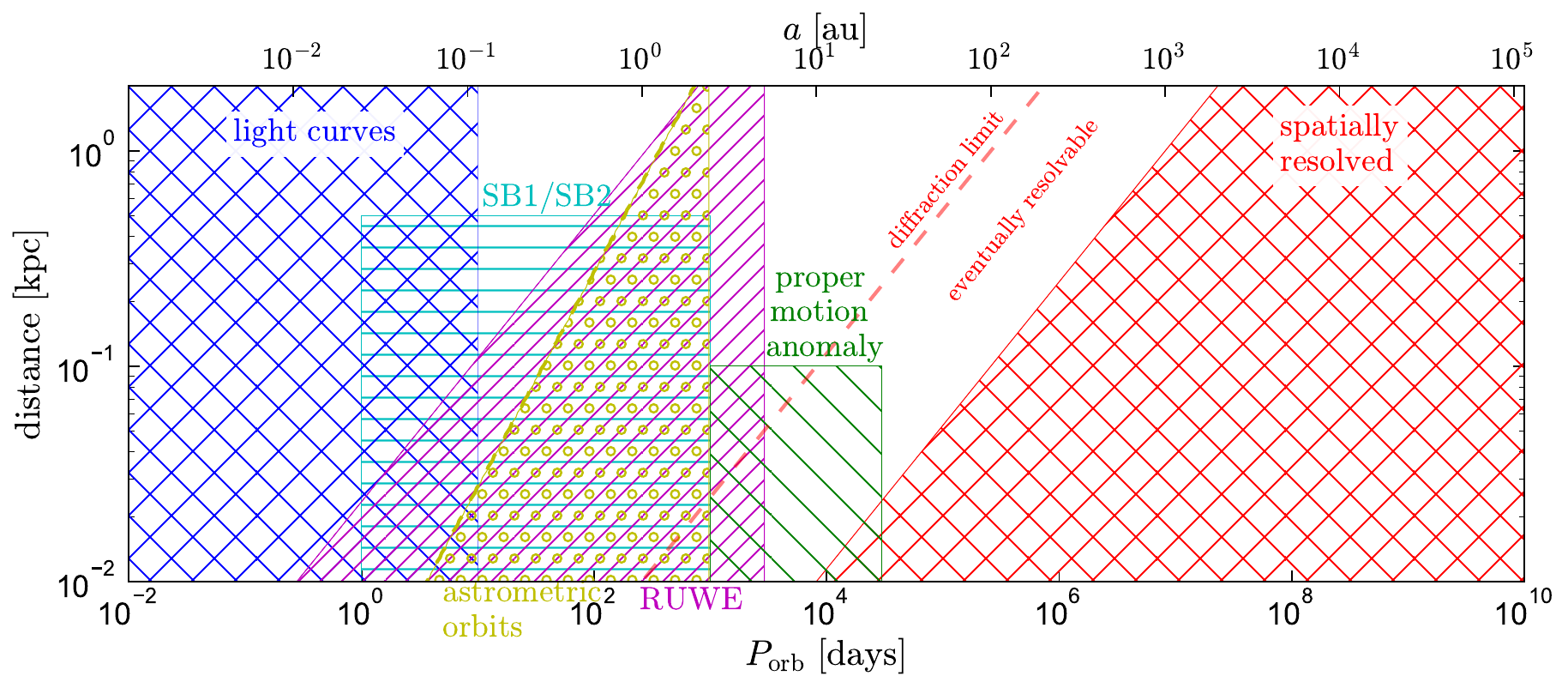}
    \caption{The binary parameter space probed by {\it Gaia} as of the mission's 3rd data release.  Hatched regions compare the approximate parameter space in which binaries are detectable via eclipses and ellipsoidal variations (``light curves''; blue), single- and double-lined radial velocities from {\it Gaia} RVS spectra (``SB1/SB2''; cyan), excess astrometric noise (``RUWE''; magenta), proper motion anomaly (green), astrometric orbits (yellow), and spatially resolved wide binaries (red). For concreteness, I consider binaries containing solar-type stars. Light curves are sensitive to the shortest periods, where the probability of photometric variability is highest. SB1/SB2 solutions probe orbits from $\sim 1$ to $\sim 10^3$ days. The upper limit is set by the {\it Gaia} DR3 observing baseline, and the lower limit by the high rotation velocities of stars in short-period binaries. RUWE is sensitive to binaries which have large angular photocenter wobbles and periods not much larger than the {\it Gaia} observing baseline; this leads to increased sensitivity at close distances. Sensitivity to binaries with astrometric orbits is similar, but with steeper drop-off at short periods due to the SNR cuts employed in DR3 (Equation~\ref{eq:par_over_err}). Proper motion anomaly is most sensitive to nearby systems with periods of order a decade, which accelerated between {\it Hipparcos} and {\it Gaia} observations. {\it Gaia} resolves binaries with separations greater than $\sim 1$ arcsec, so binaries resolved by {\it Gaia} are essentially always too wide to have interacted. {\it Gaia's} $\sim 0.1$ arcsec diffraction limit is labeled. Binaries wider than this (between the dashed red line and red hatched region) are usually not resolved in DR3, but should ultimately be resolvable with end-of-mission data.  }
    \label{fig:detection_methods}
\end{figure*}

Binaries exist over an enormous range of physical scales. They are observed in the Milky Way with orbital periods ranging from 5 minutes \citep{Roelofs2010, Burdge2020} to 100 Myr \citep{Eggleton1989, El-Badry2021_million} -- a dynamical range $\sim 10^9$ in separation and $\sim 10^{13}$ in period. Their mass ratios range from $\sim 10^{-6}$ (planets) to $\sim 10^2$ (black holes). A battery of different observational techniques is required to study binaries across this parameter space. {\it Gaia} is sensitive to essentially all possible separations, but the effective survey volume varies enormously with separation (Figure~\ref{fig:detection_methods}).

{\it Gaia} is a survey mission, whose goal is to collect a relatively small amount of data for a large number of sources. This makes {\it Gaia} well-suited for tasks such as constructing large samples, identifying rare objects, and characterizing subtle features in binary population demographics. There are also questions that can be better answered by datasets predating {\it Gaia}. For example, studies of the absolute binary fraction and period distribution are best carried out with samples with high completeness, which  require long observational baselines and a diversity of observation techniques \citep[e.g.][]{Duquennoy1991, Raghavan2010}. Studies of eclipsing and ellipsoidal binaries benefit from a large number of photometric epochs, which can currently best be provided by ground-based surveys such as OGLE \citep{Soszynski2016}, ASAS-SN \citep{Rowan2022, Li2024}, and ZTF \citep{Chen2020}. {\it Gaia's} greatest strength relative to other missions is astrometry. Even for binary populations best identified with other methods, astrometric distance constraints from {\it Gaia} have been critical for defining samples and measuring physical parameters (see Section~\ref{sec:cvs}).

We now turn to {\it Gaia's} impact on binary studies across the parameter space shown in Figure~\ref{fig:detection_methods}.

\begin{figure}[!ht]
	\centering 
	\includegraphics[width=\columnwidth]{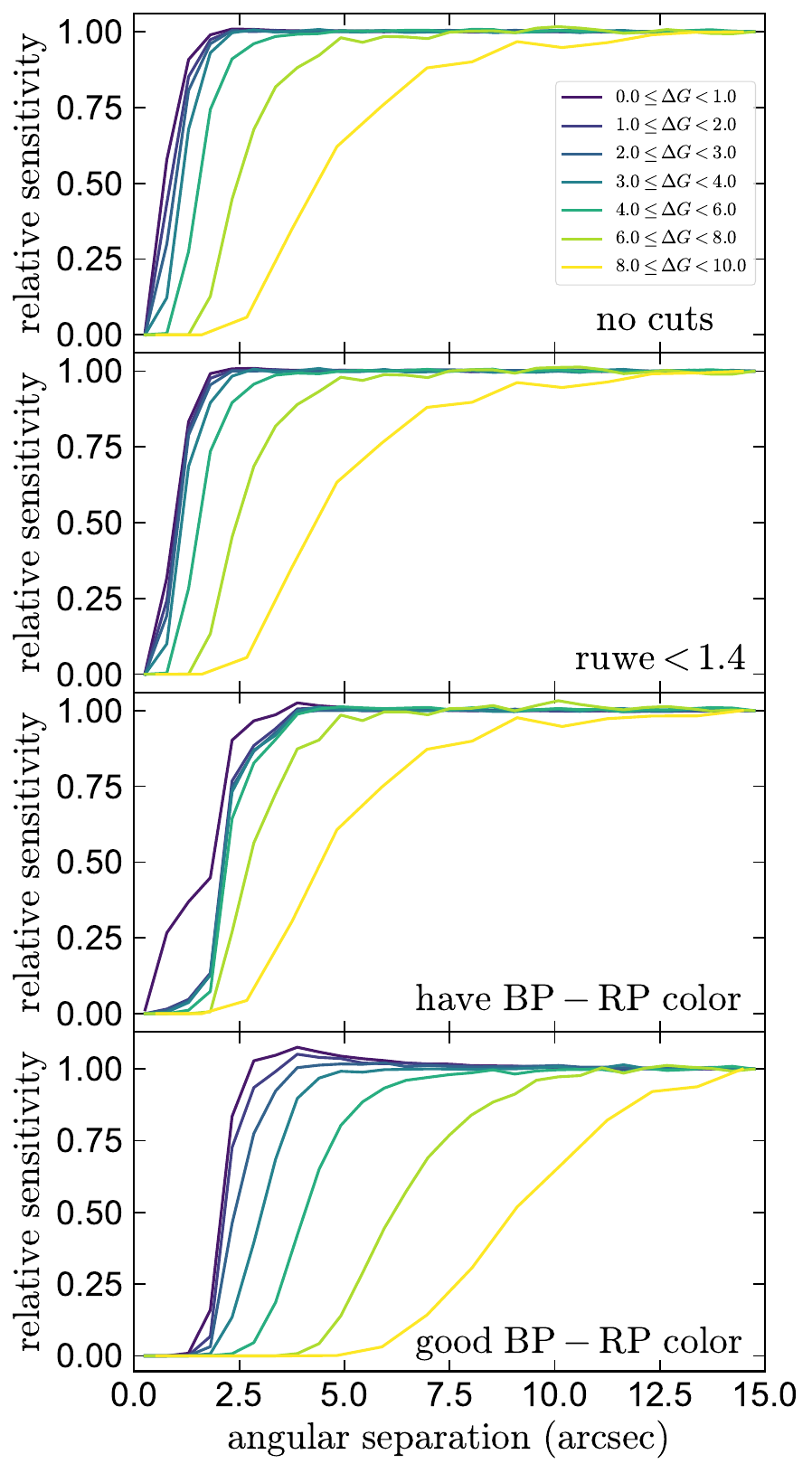}	
	\caption{{\it Gaia's} contrast sensitivity; i.e., the fraction of companions detected at a given separation,  relative to the number of similar sources that would be detected in isolation. Different lines show pairs of different magnitude difference, $\Delta G=\left|G_{1}-G_{2}\right|$, with the bluest lines showing nearly equal-brightness pairs and yellow lines showing pairs in which one component is much brighter than the other. For similar-brightness pairs, {\it Gaia's} resolution is better than 1 arcsec, with $\approx 50\%$ of pairs detected at 0.6 arcsec. Due to blending, the requirement of good \texttt{bp\_rp} colors or good astrometry for both components results in lower sensitivity. Faint companions are particularly sensitive to blending, with contaminated photometry out to $10-15$ arcsec. } 
	\label{fig:contrast}%
\end{figure}

\section{Spatially resolved binaries}
\label{sec:wide_binaries}

The widest binaries can be spatially resolved by {\it Gaia}, with single-star astrometric solutions produced for both components. {\it Gaia}'s effective angular resolution of $\sim$1 arcsec (a more precise value will be inferred below) corresponds to a minimum projected physical separation 
\begin{equation}
    s\gtrsim1000\,{\rm au}\left(\frac{d}{1\,{\rm kpc}}\right)
\end{equation}
for a binary at a distance $d$. Given that stars expand to maximum dimensions of order $\lesssim 10$\,au during their evolution,  essentially all the binaries {\it Gaia} can resolve are effectively single; i.e.,  no mass transfer between the stars is expected, and their evolution should not be altered by the presence of the companion. Such wide companions are relatively common: companions with separations $s\gtrsim 100$\,au are found around more than 10\% of solar-type stars \citep[e.g.][]{GaiaCollaboration2021gcns}.

Figure~\ref{fig:contrast} quantifies the angular resolution and contrast sensitivity of {\it Gaia} DR3. Each panel corresponds to a different set of quality cuts, with the top panel showing all sources in the catalog, the 2nd panel showing sources with \texttt{ruwe} $<1.4$ (indicative of an unproblematic astrometric solution; Section~\ref{sec:ruwe}), the 3rd panel showing sources with a published \texttt{bp\_rp} color, and the 4th sources with colors passing the quality cuts on \texttt{phot\_bp\_rp\_excess\_factor} described by \citet{Riello2021}. In all panels, the sensitivity is relative to the sensitivity at asymptotically wide separations, calculated from the two-point correlation function of chance alignments following \citet{El-Badry2018_imprints}. For near equal-brightness pairs with no quality cuts, the completeness drops to 50\% at $\approx 0.6$ arcsec, and to 0 at 0.3\,arcsec. The requirement of a good \texttt{bp\_rp} color results in significantly poorer sensitivity to close companions, since the fluxes in the BP and RP bands are dispersed over a $2\times 3$ arcsec window and thus are quite vulnerable to blending by a companion \citep[e.g.][]{Evans2018}. This means, for example, that white dwarf companions to main sequence stars are only detectable at relatively wide separations. 

Although {\it Gaia} currently resolves very few binaries with separations below 0.5 arcsec, the telescope's $\sim 0.1$ arcsec diffraction limit will ultimately allow closer pairs to be resolved. For example, {\it Hipparcos} resolved pairs with separations as small as 0.1 arcsec, about 3 times smaller than its diffraction limit \citep{Lindegren1997}. Specialized processing of astrometry for binaries would be required to achieve comparable results with {\it Gaia}. By the end of the mission, {\it Gaia} is expected to resolve equal-brightness binaries down to separations of 0.1 arcsec \citep{Harrison2023}, although it may not derive robust astrometric solutions for both components.   

\subsection{Population demographics}
Population demographics of wide binaries -- i.e., their distributions of orbital separation, mass ratio, eccentricity, metallicity, etc., -- are of interest as an observable outcome of the star formation process that can be compared to models of star and binary formation \citep[e.g.][]{Bate1997, Bate2009, Marks2012, Bate2019, Rozner2023, Guszejnov2023}.

Before {\it Gaia}, it was difficult to distinguish physically associated binaries from random chance alignments. Precise distances and proper motions from {\it Gaia} have made distinguishing binaries from chance alignments trivial since DR2 for most nearby ($d\lesssim 1$\,kpc) binaries. Only for very wide separations ($s\gtrsim 50,000$\,au), larger distances, and fainter stars, does ambiguity remain common. Figure~\ref{fig:wbs} highlights a few results from studies of wide binaries with {\it Gaia}.

\subsubsection{The separation distribution}
Catalogs of candidate wide binaries and co-moving pairs were constructed from the TYCHO-{\it Gaia} astrometric solution \citep[TGAS;][]{Michalik2015} published in {\it Gaia} DR1 by several groups. % \citep{Oelkers2017, Oh2017, Andrews2017}. 
\citet{Oelkers2017} and \citet{Oh2017} presented samples extending to separations of $\sim 3$ and $10$\,pc. \citet{Oelkers2017} found a bimodal separation distribution separated by a valley at $\sim 0.1$ pc. Similar separation distributions had been found in some earlier samples \citep[e.g.][]{Dhital2010} and were interpreted as evidence of different formation processes for binaries with $a\gtrsim 10^4$ au \citep[e.g.][]{Kouwenhoven2010}. \citet{Oh2017} found a separation distribution that increased monotonically toward wider separations and interpreted the wider pairs as ``moving groups''; i.e., associations of stars that formed together and are still slowly drifting apart, despite no longer being gravitationally bound \citep[see also][who reported similar results based on {\it Hipparcos} data]{Shaya2011}. 

\begin{figure*}[!ht]
    \centering
    \includegraphics[width=\textwidth]{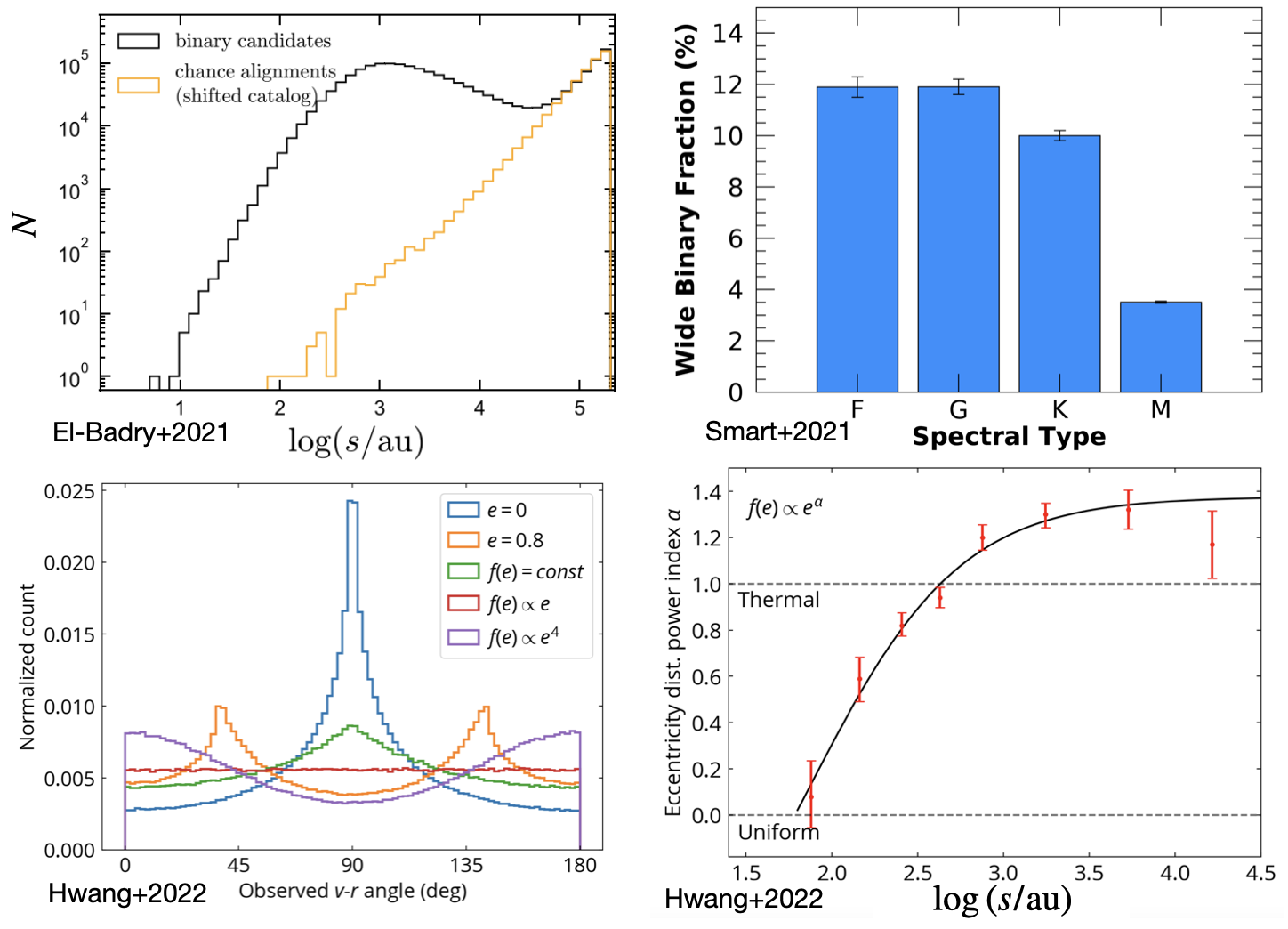}
    \caption{Wide binary population studies with {\it Gaia}. Top left panel, adapted from \citet{El-Badry2021_million}, shows the separation distribution of a sample of $\sim 2$ million wide binary candidates within 1 kpc. Gold histogram shows the expected contamination rate from chance alignments, which dominate at $s\gtrsim 30,000$ au. Top right panel, from \citet{GaiaCollaboration2021gcns}, shows the wide binary fraction as a function of spectral type within 100\,pc; this is roughly the occurence rate of companions wider than 100 au. Bottom panels, from \citet{Hwang2022_eccall}, show measurement of wide binary eccentricities. Bottom left shows the distribution of the v-r angle for simulated binaries with a range of eccentricity distributions. Bottom right shows the inferred eccentricity distribution exponent as a function of separation, as measured from the \citet{El-Badry2021_million} sample.  }
    \label{fig:wbs}
\end{figure*}

Subsequently, \citet{Andrews2017} selected a wide binary candidate sample from TGAS with an emphasis on purity, using a Bayesian method to calculate the probability that a given pair of stars is gravitationally bound. They demonstrated that the {\it Gaia} data are sufficiently precise that orbital motion within wide binaries is often detectable, even at separations of thousands of au. This means that it is generally not a good assumption that the space velocities and/or proper motions of wide binary components should be consistent within their uncertainties. \citet{Andrews2017} investigated how the separation of their candidate pairs varied with distance and astrometric uncertainties. Their selection also yielded a bimodal separation distribution when they employed less stringent cuts on distance and astrometric uncertainty. However, the relative normalization of the wide peak declined at closer distances from the Sun and essentially vanished when the strictest cuts on parallax and proper motion uncertainties were used. This led \citet{Andrews2017} to conclude that most of the pairs with separations $\gtrsim 10^5$ au reported in other works were chance alignments, not gravitationally bound binaries. Their analysis showed that the population statistics of the wide peak are consistent with what is expected for chance alignments (see e.g., the upper left panel of Figure~\ref{fig:wbs}). 

It is possible that some unbound pairs are physically associated groups of stars that formed together and are drifting apart. However, such groups are expected to be short lived: for a typical velocity dispersion of $1\,\rm km\,s^{-1}$, a pair of initially co-located stars will reach a separation $\gtrsim 1$\,pc within 1\,Myr. Dissolving star clusters may yield pairs with separations of 10-100\,pc (``moving groups''), but pairs with projected separations of $\lesssim 1$ pc can be rather cleanly separated into bound binaries and chance alignments. The division between the two groups becomes increasingly clear as the data quality improves. 

{\it Gaia} DR2 and DR3 provided significantly improved astrometric data quality, and several groups used the data to construct samples of wide binaries \citep{El-Badry2018_imprints, Jiminez2019, Hartman2020, Tian2020, Hwang2021, El-Badry2021_million}. None of these works found evidence for a bimodal separation distribution, instead finding that the separation distribution declines as ${\rm d}N/{\rm d}s\propto s^{-\gamma}$ at wide separations, with $\gamma \approx 1.6\pm 0.1$. Some additional results from these binary samples are highlighted below.

\subsubsection{Wide white dwarf binaries}
\label{sec:wb_wide}
About 16,000 resolved white dwarf (WD) + main sequence (MS) binaries, and 1,500 WD+WD wide binaries have been identified from {\it Gaia} data \citep{El-Badry2021_million, Heintz2022}. One resolved WD+WD+WD triple has also been identified \citep{Perpinya-Valles2019}. Because the ages of WDs can typically be measured with higher precision than the ages of MS stars, wide binaries containing a white dwarf provide an opportunity to age-date the companion, under the assumption that the two stars are coeval \citep[e.g.][]{Fouesneau2019, Qiu2021, Martin2021, Moss2022}. The same assumption allowed \citet{Hollands2024} to constrain the initial-final mass relation for WDs by modeling a large sample of resolved WD+WD binaries for which they obtained follow-up spectroscopy. 

\citet{Heintz2022} tested the robustness of WD age constraints by comparing the independently-measured ages of individual WDs in WD+WD binaries. They found generally satisfactory agreement for a majority of systems, with typical uncertainties of order 25\%. However, they found that 21-36\% of binaries in their sample contain a more massive WD with a shorter cooling age than its less massive companion, which is not expected if the components are coeval and the initial-final mass relation is monotonic. They attributed these discrepant ages mainly to triple evolution, wherein a merger of an inner binary can produce a massive white dwarf with a long pre-WD lifetime.

\citet{El-Badry2018_imprints} reported that the separation distribution of WD + MS and WD + WD binaries from {\it Gaia} DR2 drops off more steeply at wide separations ($s\gtrsim 3,000$\,au) than the separation distribution of MS binaries from which these systems form. They conjectured that this is a result of impulsive or asymmetric mass loss during the AGB phase, which could impart weak ($\lesssim 0.75\,\rm km\,s^{-1}$) kicks on the nascent WDs and unbind the widest binaries. \citet{Torres2022} found a similar but weaker break in the separation distribution from {\it Gaia} DR3, and \citet{Shariat2023} found that kicks were required to explain the observed separation distribution of {\it Gaia} WD triples. The physical mechanism that may produce these kicks is uncertain. 

\citet{O'Brien2024} constructed a nearly complete {\it Gaia} sample of WDs within 40 pc. The sample contains 97 resolved WD+MS binaries, and 15 resolved WD+WD binaries. The number of WD+WD binaries is at least a factor of 10 lower than predicted by binary population synthesis models \citep{Toonen2017}. This mismatch could be partially explained by WD kicks, which would unbind many would-be WD+WD binaries. It may also owe in part to the mass ratio distribution adopted in population synthesis models, which is more top-heavy than observed \citep[e.g.][]{ElBadry_2019_twin}.

\citet{Noor2024} found that the wide binary fraction of polluted DAZ WDs is similar to that of normal field WDs, suggesting that WD pollution is driven mainly by planets and is not driven primarily by binary companions, as had been proposed \citep{Bonsor2015}. Curiously, they found that DZ WDs (polluted WDs with helium-dominated atmospheres) have a wide binary fraction at least 3 times lower than DAZ or normal field WDs. Such a difference is not expected in the standard evolutionary scenarios invoked for WD pollution and spectral type evolution.

\subsubsection{The separation distribution across Galactic populations and constraints on perturbers}
\label{sec:wb_binaries}
\citet{Tian2020} explored how the separation distribution of wide binaries varies across Galactic stellar populations. It has long been appreciated that wide binaries are fragile and can be disrupted by encounters with other massive objects \citep[e.g.][]{Bahcall1985, Yoo2004, Ramirez2023}, making their separation distribution a sensitive probe of the Galactic tidal field and of small-scale structure in the gravitational potential. One might therefore expect that the separation distribution of binaries in the disk -- which suffer frequent interactions with passing stars, molecular clouds, compact objects, etc., -- would fall off more steeply toward wide separations than that of binaries in the Galactic halo. 

\citet{Tian2020} tested this hypothesis using wide binaries from {\it Gaia} DR2. They found that kinematically selected samples of binaries on halo-like and disk-like orbits have essentially identical intrinsic separation distributions out to separations of at least $10^4$ au. At the widest separations, they found modest evidence for a {\it steeper} decline in the separation distribution of halo binaries than in that of disk binaries, contrary to the trend that might be expected due to disruption of wide binaries by dynamical encounters. \citet{Tian2020} speculated that this might reflect the different star formation conditions from which the halo and disk binaries were born. In any case, this work showed that if precise measurements of dynamical perturbations are sought from wide binaries, then the {\it initial} separation distribution of wide binaries is a dominant source of uncertainty.

\subsubsection{Eccentricities}
\label{sec:eccentricity}
Essentially all binaries that are spatially resolved by {\it Gaia} have orbital periods that are much too long for significant orbital motion to be observed on a timescale of a few years. This makes eccentricity measurements for individual wide binaries impossible. However, the {\it distribution} of wide binary eccentricities can still be studied from the distribution of the relative proper motions of the components (both speed and direction), which probe the orbital velocity at a particular snapshot in time. A few approaches have been used to exploit this. 

Using Monte Carlo simulations, \citet{Tokovinin2020} applied the method developed by \citet{Tokovinin2016} to a sample of $\sim 2600$ binaries from {\it Gaia} DR2 within 67 pc of the Sun. This approach forward-models the distribution of the angle, $\gamma$, between a binary's position angle and relative velocity vector, and the amplitude of the projected orbital speed normalized by separation and the binary's total mass. Limiting his analysis to low-mass binaries ($M_1 \lesssim 1\,M_{\odot}$) and attempting to remove higher-order multiples, he found that the eccentricity distribution of the full sample was close to thermal ($f(e)\,{\rm d}e =2e\,{\rm d}e$, corresponding to a median eccentricity of about 0.71). He also found a relative dearth of highly eccentric binaries at separations less than 200 au (a sub-thermal distribution, but still more eccentric than a uniform distribution), and a slightly super-thermal distribution at separations wider than 1000\,au.

\citet{Hwang2022_eccall} used a related method to study the eccentricity distribution of a larger sample of $\sim 10^5$ wide binaries within 200 pc from {\it Gaia} DR3. Using the method originally developed by \citet{Tokovinin1998} -- which depends only on the distribution of the $\gamma$ angle and not on the magnitude of the orbital speed -- they inferred the eccentricity distribution as a function of separation. Assuming binaries are randomly oriented and observed at random phases, one expects circular orbits to preferentially be observed with $\gamma = 90$ deg (perpendicular position and relative velocity vectors), while eccentric orbits should preferentially have $\gamma = 0$ deg and $\gamma = 180$ deg (aligned position and velocity vectors). This is illustrated in the bottom left panel of Figure~\ref{fig:wbs}. While this method loses some information compared to the approach of \citet{Tokovinin2020}, it has the advantage of not requiring estimates of stellar masses and is likely less sensitive to the effects of unrecognized higher-order multiples. \citet{Hwang2022_eccall} inferred an eccentricity distribution consistent with uniform at separations of $\lesssim 100\,{\rm au}$, and steadily increasing eccentricities over $100-1000$\,au (Figure~\ref{fig:wbs}, bottom right). At separations $a\gtrsim 1000$\,au, they found a super-thermal eccentricity distribution, $f(e)\propto e^{1.3}$, corresponding to a median eccentricity of 0.74, and with 22\% of binaries having $e > 0.9$. They also found modest evidence for a shallower eccentricity distribution among wide binaries hosting inner subsystems. 

The results of the studies by \citet{Tokovinin2020} and \citet{Hwang2022_eccall} are, broadly speaking, consistent: both studies find increasingly eccentric orbits at wider separations and a super-thermal eccentricity distribution at the widest separations. Several models have been proposed to interpret these observations \citep[e.g.][]{Hamilton2022, Grishin2022, Xu2023,  Hamilton2023, Rozner2023}.

\subsubsection{The mass ratio distribution and twin binaries}
\label{sec:twin}
The homogeneity and well-understood selection function of the {\it Gaia} data allowed \citet{ElBadry_2019_twin} to infer the intrinsic mass ratio distribution of wide binaries as a function of primary mass and separation. At all separations and masses, the mass ratio distribution is inconsistent with random pairing from the IMF, favoring higher mass ratios than expected in this scenario.  An unexpected result of this analysis was the discovery of a population of equal-mass ``twin'' binaries with mass ratios $q\gtrsim 0.95$ extending to separations of several hundred to several thousand au, depending on primary mass. 

The presence of a twin binary excess was well-documented among close binaries with separations $\lesssim 0.1\,\rm au$ \citep{Tokovinin2000}, where it had been ascribed to formation processes that equalize the mass ratio, such as accretion from a circumbinary disk. If the same mechanism is to form twin binaries at wide separations, some mechanism is required to widen the binaries after formation, since the twin excess extends to separations wider than the largest observed circumbinary disks. Using speckle interferometry, \citet{Tokovinin2023} showed that an excess of twins is also present at 10-100\,au, below the {\it Gaia} resolution limit.

\citet{Hwang2022} showed that the eccentricities of the wide twins are on average higher than those of non-twins with similar separations. That is, although twins today are observed with separations of hundreds to thousands of au, they generally reach separations of $10-100$\,au at periastron. This observation lends credence to the hypothesis that twins formed via accretion from circumbinary disks, but further modeling is needed to work out the details of how their orbits were subsequently perturbed. 

\subsubsection{Planets}
Another possible manifestation of correlated formation of wide binaries can be found in their planetary systems: several studies have reported an excess of planets in wide binaries whose orbits are aligned with the orbital plane of the wide binary \citep{Dupuy2022, Christian2022} and also with the spin axis of their host stars \citep{Rice2024}. This excess appears to extend to binary separations of at least several hundred au. Like the excess of twin wide binaries, this observation seems to require ``spooky action at a distance'' on scales greater than typically associated with accretion processes during star formation. 

\subsubsection{Metallicity dependence}
\label{sec:metallicity}
Variation of the binary fraction with metallicity has been a subject of many investigations over the last several decades, with many conflicting results in the literature. In a meta-analysis of several different binary samples, \citet{Moe2019} found that when selection effects are accounted for, the {\it close} binary fraction of solar-type stars is strongly anti-correlated with metallicity, while the trend disappears at wide separations. Correlations with metallicity for massive stars were found to be weaker. The boundary between close and wide binaries was fuzzy but was proposed to correspond to the transition between the disk fragmentation and core fragmentation formation mechanisms. 

\citet{El-Badry2019_feh} attempted to constrain the separation below which metallicity dependence appears using a sample of wide binaries from {\it Gaia} DR2 and metallicities from several wide-field surveys. They found that the wide binary fraction is independent of metallicity at $s\gtrsim 250$\,au but becomes anticorrelated with metallicity at closer separations, with the anticorrelation at $50-100$\,au almost as strong as \citet{Moe2019} and \citet{Badenes2018} found at $<10$\,au. They thus proposed that most binaries with separations $\lesssim 100$ au form by disk fragmentation, as expected in the model advocated by \citet{Moe2019}.

 \citet{Hwang2021} and \citet{Hwang2022_feh} focused on wide binaries ($\gtrsim 1000$\,au) and found more complicated behavior than the flat trend with metallicity reported by \citet{El-Badry2019_feh}: they found that the wide binary fraction peaks at $[\rm Fe/H]= 0$ and declines towards both higher and lower metallicities. Curiously, comparison of their sample and the sample used by \citet{El-Badry2019_feh} suggested that the discrepant results are in large part due to different treatments of hierarchical triples: \citet{El-Badry2019_feh} employed cuts on relative proper motion that excluded most triples, while \citet{Hwang2022_feh} included them, and the triples seem to cluster at $[\rm Fe/H]=0$. \citet{Niu2022} studied the metallicity dependence of the wide binary fraction further using metallicities from the LAMOST survey. They also found an anticorrelation with metallicity at $s < 200$\,au, and a more complicated trend that depends on stellar mass and age at wider separations. 

\subsection{Triples and quadruples}
\label{sec:higher_order}

\begin{figure}[!ht]
	\centering 
	\includegraphics[width=\columnwidth]{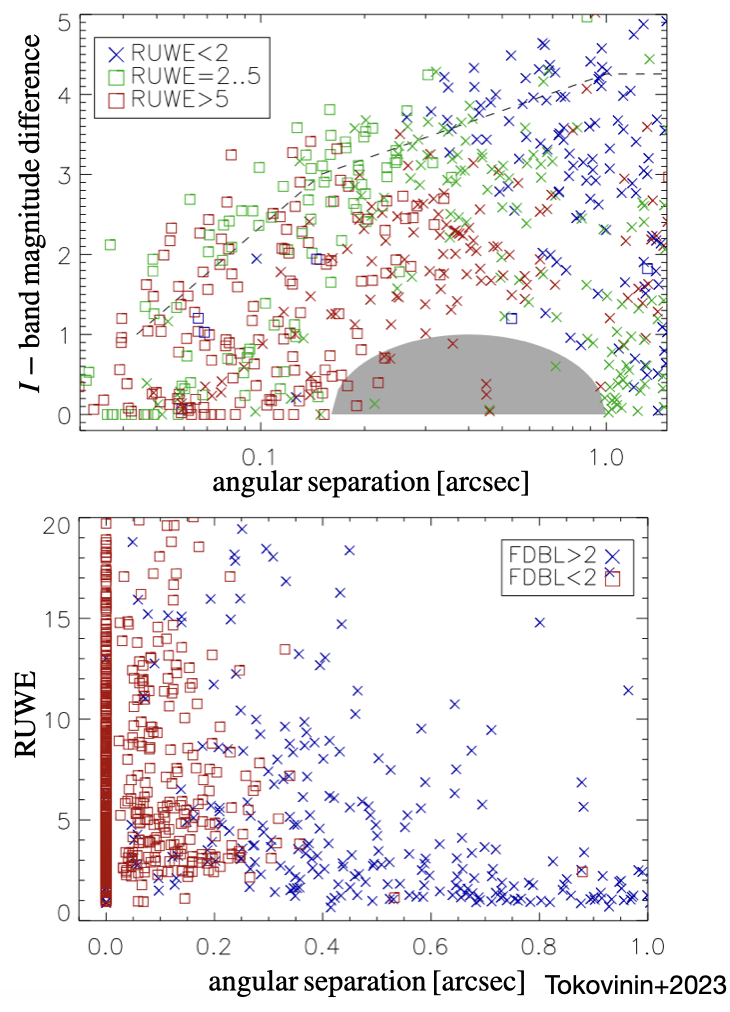}	
	\caption{Hierarchical triples discovered by \citet{Tokovinin2023} through speckle-imaging follow-up of {\it Gaia} wide binaries. Most of the inner binaries were too tight to be resolved by {\it Gaia}, but the {\it Gaia} quality flags \texttt{ruwe} and \texttt{ipd\_frac\_multi\_peak} (referred to as FDBL in the figure) hinted at their presence. Dashed black line in the top panel shows the speckle sensitivity limit. Shaded gray region shows that sources with a comparably bright companion that is marginally resolved by {\it Gaia} do not enter the sample in the first place, because most of these only received 2-parameter solutions in DR3. Most companions with separations greater than $\approx 0.2$ arcsec have \texttt{ipd\_frac\_multi\_peak} $> 2$. In this (relativley nearby) sample, most sources with a companion closer than $\approx 0.5$ arcsec have \texttt{RUWE} $>2$.}
	\label{fig:speckle}%
\end{figure}

A significant fraction of the wide binaries discovered by {\it Gaia} are actually hierarchical triples and higher-order multiples in which close companions to one or both components are unresolved by {\it Gaia}. A range of methods can be used to characterize these inner binaries, some of which are discussed below. For a recent review of higher-order multiplicity, I direct the reader to \citet{Tokovinin2021}.

\subsubsection{High-resolution imaging}
\label{sec:speckle}
\citet{Tokovinin2023} carried out an extensive speckle interferometry campaign with the high-resolution camera on the 4.1m SOAR telescope focused on resolving subsystems within wide binaries identified by {\it Gaia}. Targets were selected from the 100 pc sample of \citet{GaiaCollaboration2021gcns} based on their having a large \texttt{ruwe} (section~\ref{sec:ruwe}), a large \texttt{ipd\_frac\_multi\_peak} (meaning that two peaks were detected in a significant fraction of transits), or a large RV uncertainty (potentially indicative of RV variability). Observations of 1243 candidates resulted in 503 new subsystems being resolved (Figure~\ref{fig:speckle}). The newly resolved subsystems have separations ranging from 0.03 to 1 arcsec, peaking at about 0.1 arcsec. Speckle observations are thus sensitive to companions about a factor of 10 closer than the {\it Gaia} resolution limit. Since the separation distribution of wide binaries rises toward close separations, peaking at 10-50\,au, this represents a significant improvement compared to what can be achieved with {\it Gaia} alone. Most of the hierarchies identified by \citet{Tokovinin2023} fall in the dynamically stable regime with $P_{\rm outer}/P_{\rm inner} \gtrsim 5$. Beyond this requirement of dynamical stability, the separation distribution of the outer orbits appears close to the average for the field binary population. Joint analysis of the {\it Gaia}-resolved wide binaries with and speckle-resolved tighter binaries within the 100 pc sample reveals a trend of increasingly unequal mass ratios at wider separations. 

Among other applications, the speckle sample makes it possible to assess how different {\it Gaia} flags and quality metrics are sensitive to binarity (Figure~\ref{fig:speckle}). Most pairs with separations wider than $\sim$0.2 arcsec have \texttt{ipd\_frac\_multi\_peak} $>$ 0, meaning that {\it Gaia} detected more than one peak when the source transited the focal plane with at least some scan angles. This implies that \texttt{ipd\_frac\_multi\_peak} is a useful diagnostic for detecting marginally companions, with the caveat that resolved companions within $\approx 2.5$ arcsec can also lead to multiple peaks, which largely be separated from unresolved companions in the separation vs. \texttt{ipd\_frac\_multi\_peak} plane.

Many stars with close companions also have large \texttt{ruwe} values, indicating a problematic astrometric solution. For the binaries in the \citet{Tokovinin2023} sample that are closest to Earth, this is likely a consequence of orbital motion, which causes photocenter wobble that cannot be well-fit by a single-star astrometric model. However, the majority of the sources with elevated \texttt{ruwe} values have separations of $>10$\,au, corresponding to orbital periods of $\gtrsim 30$\,yr. At these periods, orbital motion is largely expected to be absorbed into proper motion, and a single-star model should provide an acceptable fit to the astrometric motion of either component, or that of the photocenter. I thus conclude that enhanced \texttt{ruwe} value can result either from photocenter acceleration in binaries with periods of a few years -- corresponding to angular separations of $\rho\sim0.03\,{\rm arcsec}\left(\frac{\varpi}{10\,{\rm mas}}\right)$  -- {\it or} from wider pairs in which a marginally resolved companion simply disturbed the {\it Gaia} astrometric measurements. Only at close distances ($d\lesssim 30$ pc) are typical binaries expected to be affected by both issues. In many cases, the \texttt{ipd\_frac\_multi\_peak} parameter can distinguish between the two situations: sources with \texttt{ipd\_frac\_multi\_peak} $> 0$ are likely marginally resolved pairs and likely to have problematic astrometric measurements, while being too wide for their physical acceleration to be detectable.

Another lesson from speckle follow-up is that in resolved pairs with separations closer than about 1 arcsec, {\it Gaia} often provides only a 2-parameter astrometric solution for the fainter star, or for both stars if the components have similar brightness. This means that a significant fraction of resolved wide binaries will be missed by {\it Gaia} searches which require both components to have similar proper motions and/or parallaxes. Some close pairs will of course simply be chance alignments, but the rate of chance alignments among bright stars in sparse fields is low enough that selections can be optimized to find true pairs \citep[see e.g.][for a recent demonstration]{Medan2023}.

\subsubsection{Larger-than-expected proper motion difference}
\label{sec:dv}
The two components of a wide binary are expected to have nearly identical proper motions, since the typical orbital velocity for solar-type binaries only allows for a velocity difference of order $\Delta v\sim1\,{\rm km\,s^{-1}}\left(\frac{s}{1000\,{\rm au}}\right)^{-1/2}$. But if one component has a close unresolved companion, this can induce an additional plane-of-the sky velocity of up to several $\rm km\,s^{-1}$ in that component. In many cases, this will make the wide binary appear to be unbound. Triples can thus be identified as wide binaries in which the apparent velocity difference between the two resolved components is larger than expected for a bound orbit, but still smaller than expected for chance alignments \citep[e.g.][]{El-Badry2021_million}. Conversely, the fraction of triples included in a wide binary sample will be lower for samples imposing a Keplerian cut on the proper motion difference \citep[see][]{Tokovinin2023}.

\subsubsection{Photometric variability}
\label{sec:photo_variability}
Hierarchical multiples hosting a tight subsystem have also been identified by searches for resolved wide binaries in which one  component is an eclipsing binary or ellipsoidal variable. A systematic effort was undertaken by \citet{Hwang2020}, who identified a sample of 1333 candidate contact binaries via {\it Gaia} light curve variability and then used {\it Gaia} astrometry to compare the occurrence rate of wide companions in this sample to the rate for a control sample of field stars. They found that $\approx 14.1\pm 1\%$ of contact binaries have a resolved wide companion with separation $10^3-10^4$\,au, compared to only $4.5\pm 0.6\%$ of field stars in their control sample. Their search was sensitive to main-sequence companions with $M\gtrsim 0.2\,M_\odot$. This factor-of-3 enhancement in the rate of wide companions suggests that wide companions play an important role in the formation of close binaries.  Similarly, \citet{El-Badry2022_magnetic} found that the occurrence rate of companions with $s>2000$\,au is 15\% for eclipsing M dwarf binaries from ZTF, compared to only $\approx 1.3$\% for a control sample of M dwarfs. These occurrence rates are lower limits, since many wide companions will be too close to be resolved by {\it Gaia}. Detecting all such companions is a formidable task, even for controlled samples in a small volume. In his now-seminal study using adaptive optics imaging of nearby spectroscopic binaries, \citet{Tokovinin2006} found that $96\pm 7\%$ of solar-type binaries with $P<3$\,d have a wide companion. $> 80\%$ of these tertiaries were too close to enter the samples from \citet{Hwang2022} and \citet{El-Badry2022_magnetic} discussed above.

\citet{Fezenko2022} searched for eclipsing binaries within resolved wide binaries and identified 8 such 2+2 quadruples. This was a factor of $\sim7$ more than they would have expected to find if wide binaries were formed by randomly pairing field stars. That is, if a star is an eclipsing binary, it is more likely that its wide companion is an eclipsing binary than if it is not. They interpreted this finding as resulting partly from the age- and metallicity- dependence of the binary fraction (since the components of wide binaries are generally coeval and have the same metallicity), and partly reflecting the fact that wide tertiaries aid the formation of tight binaries.

\subsection{Tests of gravity}
\label{sec:tests_gravity}
Binaries with separations larger than $\sim 10^4$ au have sufficiently low internal accelerations that they are in the so-called ``deep MOND'' regime: that is, the Modified Newtonian Dynamics \citep[MOND;][]{Milgrom1983} theory of gravity predicts measurably different orbital velocities from general relativity. Several studies have thus used observations of wide binaries in attempts to measure the gravitational force law. Such measurements have the potential to test modified gravity theories while avoiding the various complications associated with testing these theories on galactic scales \citep[see e.g.,][]{Famaey2012}.

The basic measurement is straightforward: at fixed total mass, MOND predicts a higher orbital velocity than Newtonian gravity. The 3D velocity cannot be measured with high precision with {\it Gaia} alone, but the projected plane-of-the-sky orbital velocity can be precisely constrained from the difference in the components' proper motions \citep{Banik2018, Banik2019}.

Several factors complicate this measurement. In the Galactic plane, the ``external field effect'' diminishes the deviation between the MOND and Newtonian velocity predictions to only $20\%$, compared to factors of order unity for an isolated binary or one in the distant Galactic halo. Projection effects tend to increase the apparent velocity difference of wide binaries \citep{El-Badry2019_gravity}.   Unrecognized triples and higher-order multiples generically increase the apparent velocity differences of wide binaries (Section~\ref{sec:dv}). As a result, the force law inferred from measurements of binaries' projected velocity differences depends significantly on the assumed mass--luminosity relation, eccentricity distribution, and the assumed population of unresolved inner binaries. There is thus still disagreement in the literature about what {\it Gaia} wide binaries tell us about the gravitational force law. 

\citet{Hernandez2019, Hernandez2022} and \citet{Hernandez2023} reported that the projected rms velocity difference of nearby observed binaries follows a Keplerian decline at separations below $\sim 2000$ au, but that the observed velocity difference flattens off at wider separations. Such behavior is unexpected in Newtonian gravity, but also in MOND, where the external field effect leads to only a $\sim 20\%$ reduction in binaries' orbital velocities in the solar neighborhood, still with a $\Delta v\propto s^{-1/2}$ decline. 

\citet{Pittordis2019} found that the observed distribution of total mass-scaled velocity differences in wide binaries exhibits a long tail that was not predicted by simulations assuming either Newtonian gravity or MOND. It is in large part the existence of this tail that causes the flattening of the rms velocity difference highlighted by Hernandez et al. Subsequently, \citet{Clarke2020} showed that the tail can be explained as a consequence of unrecognized higher-order multiplicity (Section~\ref{sec:dv}). Higher-order multiplicity increases the predicted velocity difference between the two observed components of a wide binary for two reasons: (a) the total mass of one component is larger than inferred from a single-star mass-luminosity relation, and (b) the orbital velocity of one component of a wide binary around an unresolved faint companion typically exceeds the velocity of the pair around a wide companion. 

Several works have attempted to remove higher-order multiples from observed samples by filtering on criteria such as \texttt{ruwe}, position in the color-magnitude diagram (CMD), or RV variability. It is important to note, however, that such cuts will only catch a fraction of inner binaries. For example, for solar-type stars at a distance of $\approx 100$\,pc, at least half of all close companions are too faint to detect photometrically \citep{Riddle2015, Tokovinin2023}, while cuts on \texttt{ruwe} and related factors will be insensitive to companions with separations ranging from 5-25\,au. For these reasons, it may be better to forward-model the population of inner binaries than to attempt to remove higher-order multiples from the sample. Or, if higher-order multiples are to be removed, more stringent vetting is required than can be accomplished with {\it Gaia} data alone.

\citet{Pittordis2023} studied the pairwise relative projected velocity distribution of $\sim 70,000$ wide binaries within 300 pc of the Sun using data from {\it Gaia} eDR3. When they included a model for higher-order multiples (and also unbound ``flybys''), they found a formally highly significant preference of classical gravity over MOND. They considered modeling of the population of higher-order multiples to be the largest uncertainty in their analysis. In another study, \citet{Banik2024} found Newtonian gravity to be strongly preferred ($16\sigma$) over MOND, with relatively weak sensitivity to the assumed eccentricity distribution and parameters of the inner binary population. However, they required an inner binary fraction that is likely unphysically high to match the distribution of observed velocity differences. 

On the other hand, \citet{Chae2023} carried out a similar analysis and found that MOND was strongly preferred ($10\sigma$) over classical gravity. Despite significant discussion in the literature \citep[e.g.][]{Hernandez2023b, Banik2024, Chae2024}, the waters remain muddy. The conclusions drawn by readers about these results will likely correlate strongly with their priors. One clear takeaway is that the formal significance of statistical tests reported in the large-$N$ regime is often not particularly useful, because systematic rather than random uncertainties dominate. 

There are several possible avenues for improvement in future studies. Using speckle imaging and precision RV follow-up, a clean sample of wide binaries with few inner tertiaries could be selected \citep[e.g.][]{Manchanda2023}. Precision RV follow-up for both components of a sample of wide binaries would be useful for measuring the third (radial) component of their orbital velocity, removing possible systematics and/or underestimated uncertainties in the projected velocities inferred from {\it Gaia} proper motions. Finally, RV follow-up could be used to measure the radial velocity differences of {\it Gaia}-selected wide binaries at large distances ($d\sim 20$\,kpc) in the Galactic halo. The advantage of this approach is that the external field effect is much weaker there, and so the predicted velocities in classical gravity and MOND differ by large factors rather than by only $\approx 20\%$.

\subsection{Dynamical mass measurements}
The distribution of proper motion differences in wide binaries can also be used to empirically constrain the mass-luminosity relation. This is essentially the same approach used in the studies discussed in Section~\ref{sec:tests_gravity}, except that one assumes a fixed gravitational force law and solves for mass, instead of assuming fixed mass and solving for the gravitational force law. \citet{Giovinazzi2022} used $\sim 4000$ wide binaries with nearly equal-mass components to constrain an empirical $M_{G_{\rm RP}}-$mass relation for main-sequence stars. \citet{Hwang2024} extended this work to include unequal-mass binaries across the CMD.

Empirical mass measurements in wide binaries can also be carried out with RVs, particularly if one component is significantly more compact than the other. The difference between the RVs of the two components constrains the gravitational redshift,  $v_g \propto M/R$, of the more compact star, and the spectral energy distribution and parallax constrain the radius. This allowed \citet{El-Badry2022} to measure the masses of main-sequence stars with wide red giant companions using {\it Gaia} RVs, and \citet{Arseneau2024} to constrain the mass-radius relation of WDs in wide binaries with a main-sequence companion.

\section{Astrometric perturbations}
\label{sec:ruwe}

\begin{figure*}
    \centering
    \includegraphics[width=\textwidth]{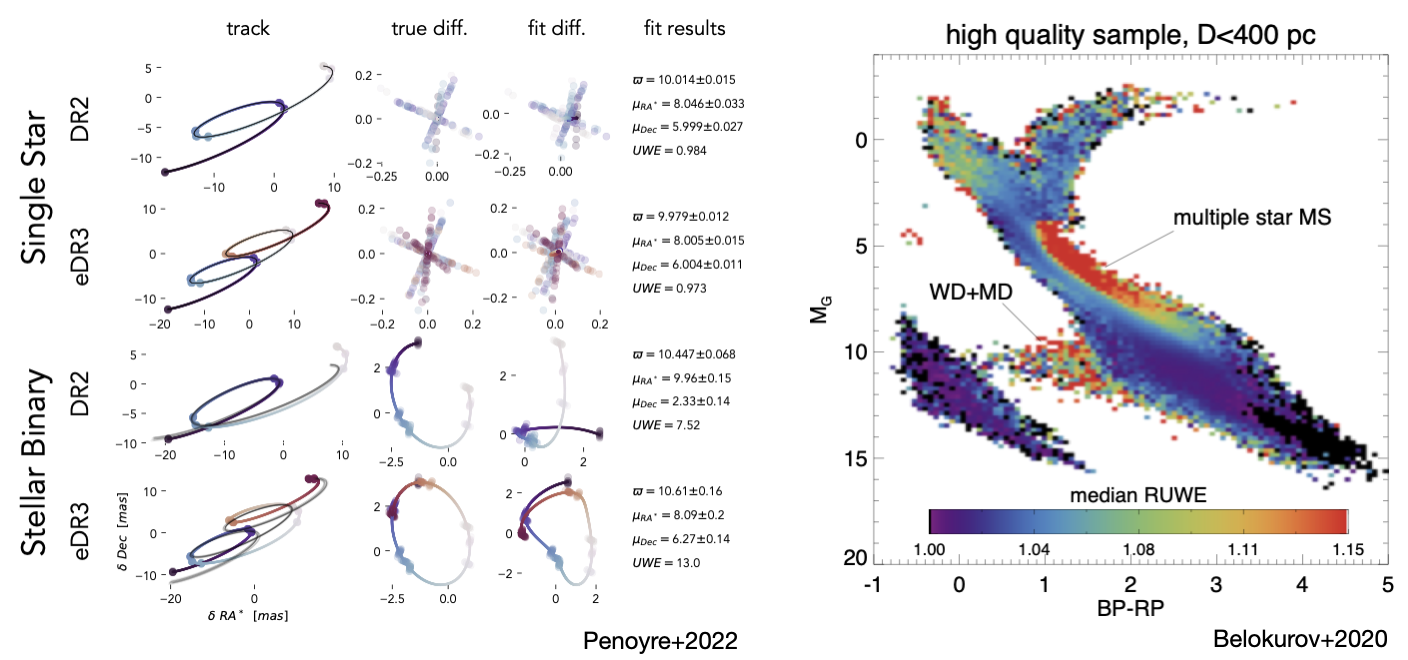}
    \caption{Astrometric noise due to binarity. Left panel, adapted from \citet{Penoyre2022}, shows simulated epoch astrometry and best-fit single-star fits for a single star (top) and an unresolved binary (bottom; with $P_{\rm orb} = 2\,{\rm yr}, e=0.5$ and $\varpi=10$\,mas). Such a binary leads to residuals $\sim 10$ times larger than expected for a single star with similar observable properties. Right panel, adapted from \cite{Belokurov2020}, shows the {\it Gaia} DR2 color-magnitude diagram of stars within 400 pc, with each pixel colored by the median \texttt{ruwe} (a measure of the astrometric residuals of a single-star fit) in that pixel. Higher than average \texttt{ruwe} values are found in regions where one expects to find large numbers of binaries: above the main sequence, in the hot subdwarf clump, and in the region between the main sequence and WD cooling track, where unresolved WD + M dwarf binaries are found. }
    \label{fig:ruwe}
\end{figure*}

\subsection{RUWE}
{\it Gaia's} standard 5-parameter astrometric model assumes that the motion of a source can be explained as a consequence of parallax and proper motion alone. This assumption in general does not hold for binaries. In some cases, binary orbits can be constrained by fitting more complicated astrometric models (Section~\ref{sec:orbits}). But binary models were only published in DR3, and only for a modest number of targets. Many binaries can nevertheless be detected -- and to some extent, characterized -- on the basis of a poor single-star astrometric model fit. This is illustrated in the left pane of Figure~\ref{fig:ruwe}, which shows that the expected residuals of a single-star fit for nearby binaries can be significantly larger than the uncertainty of the astrometric measurements. 

{\it Gaia} includes several astrometric goodness-of-fit indicators that are potentially diagnostic of binarity. The most robust of these is the \texttt{ruwe} parameter (``renormalized unit weight error''; \citealt{lindegren2018_ruwe}), which can be roughly conceptualized as the reduced $\chi^2$ of the astrometric fit, renormalized to correct from trends with apparent magnitude and color due to data processing issues.

\citet{Belokurov2020} showed that selecting targets with large \texttt{ruwe} (e.g. \texttt{ruwe} $>1.4$) yields a sample of sources preferentially in regions of the color-magnitude diagram where binaries are expected to be concentrated (Figure~\ref{fig:ruwe}, right panel). Models to predict \texttt{ruwe} for an arbitrary binary were developed by this work and by \citet{Penoyre2022, Penoyre2022b} and \citet{Andrew2022}. For binaries with orbital periods shorter than the {\it Gaia} observational baseline, the  expected amplitude of the \texttt{ruwe} signal due to an unresolved binary scales approximately linearly with the angular size of the photocenter orbit \citep[e.g.][]{Stassun2021}. As of {\it Gaia} DR3, \texttt{ruwe} is thus most sensitive to binaries that are nearby, have orbital periods of order 1000 days, and have massive but faint companions.

\citet{Penoyre2022b} constrained the unresolved binary fraction as a function of CMD position in the 100 pc sample using a re-renormalized version of \texttt{ruwe}. \citet{Korol2022} studied the \texttt{ruwe} distributions of nearby white dwarfs and found evidence for a dearth of WD+WD binaries with au-scale orbits (i.e., the separation range to which \texttt{ruwe} is most sensitive), presumably because most orbits in this separation range are cleared out by common envelope evolution. 

\citet{Andrew2022} showed that a combination of astrometric and RV excess noise can in principle be exploited to find massive companions: sources that have both large \texttt{ruwe} and significant RV variability must have massive companions, if the astrometric wobble and RV variable are both reliable and due to the same companion. Before {\it Gaia} DR3, they assembled a ``gold'' catalog of 45 compact object binary candidates, as well as 4600 lower-confidence candidates. Some of these candidates have astrometric and/or RV binary solutions in {\it Gaia} DR3 (Section~\ref{sec:orbits}). Examination of these solutions revealed that the largest contaminant for their search strategy is hierarchical triples, in which large-amplitude RV variability can be due to an inner binary, while a large orbit and \texttt{ruwe} is a result of a wider tertiary.

A few cautionary remarks on interpretation of {\it Gaia} \texttt{ruwe} values are warranted, not relating to the studies discussed above. First, \texttt{ruwe} is only sensitive to binaries with periods less than a few $\times10^3$ days and photocenter orbits that are comparable to {\it Gaia's} astrometric precision. A common mistake in the literature has been to conclude that a target is not a binary because it does not have an inflated \texttt{ruwe} value, but many -- indeed, most -- binaries will escape detection via \texttt{ruwe} because their orbits are too tight or have too long periods, because they are too far away, or because they are too faint. Conversely, a large \texttt{ruwe} value does not guarantee that a source is a binary: any problem in the astrometric solution can result in a large \texttt{ruwe} value. Perhaps the most common false-positive is a marginally resolved neighboring source (physically associated or not) that distorts the PSF (Section~\ref{sec:speckle}). 

\subsection{Proper motion anomaly}
\label{sec:prop_motion_anomaly}
Astrometric wobble due to companions on somewhat longer-period orbits can be probed using {\it proper motion anomaly} (PMa); changes in the measured proper motion of a star over time due to orbital acceleration \citep{Brandt2018, Kervella2019, Brandt2021}. The presence of a dark companion could thus be detected, for example, from the fact that a star has significantly different proper motions measured in {\it Gaia} DR2 and DR3. In practice, the most powerful PMa constraints thus far have come from comparison of {\it Gaia} proper motions with the proper motion inferred from comparison of {\it Gaia} and {\it Hipparcos} positions. Because the sensitivity of proper motion measurements increases linearly with the observational baseline at fixed number of epochs, {\it Hipparcos}-{\it Gaia} proper motion can be measured significantly more precisely than the {\it Hipparcos}-only proper motion. While accelerations between {\it Gaia} DR2 and DR3 can also be significant, in most cases the binaries detected from such accelerations will also be detectable via enhanced \texttt{ruwe}.

PMa is most sensitive to binaries with orbital periods ranging from a few years (the baseline over which {\it Hipparcos} and {\it Gaia} DR2 or DR3 measurements are averaged) to $\sim 100$ years (a few times longer than the $\sim$25 year baseline separating the {\it Hipparcos} and {\it Gaia} solutions). For shorter periods, the {\it Gaia} proper motion is an average over multiple orbits and thus loses its sensitivity. For longer periods, the orbital motion will be absorbed into the longer-term {\it Hipparcos}-{\it Gaia} proper motion. 

PMa is a measurement of a binary's time-averaged acceleration. It is particularly sensitive to low-mass companions with separations of order 10\,au, which result in periods too long to be detectable with {\it Gaia} astrometry alone and benefit from the long time baseline between the {\it Hipparcos} and {\it Gaia} solutions. The gravitational effects of the directly-imaged planet $\beta$ Pic b on its host star, for example, were detected by \citet{Kervella2019} with high significance. A small PMa also ruled out some of the long-controversial planets orbiting the post-common envelope binary HW Vir \citep{Baycroft2023}. 

Like \texttt{ruwe}, the sensitivity of PMa scales inversely with distance, making the technique most powerful for the nearest stars. On the other hand, {\it Gaia} astrometry suffers from systematics for very bright stars; indeed, most stars with good {\it Hipparcos} solutions are bright enough that their {\it Gaia} astrometry is fully systematics-limited. The sweet spot for PMa is  nearby low-mass stars.

A potential impediment to companion detection with PMa lies in mismatch between the {\it Hipparcos} and {\it Gaia} coordinate reference frames. \citet{Brandt2018} and \citet{Brandt2021} empirically calibrated to {\it Hipparcos} and {\it Gaia} astrometric frames using the ansatz that a majority of sources do not experience significant accelerations and thus should have statistically identical astrometric solutions when observed by different surveys and propagated to  a common epoch. For {\it Gaia DR2}, this analysis revealed significantly underestimated proper motion uncertainties and a global frame rotation. These systematics were significantly reduced in DR3, but \citet{Brandt2021} found that a $\approx 35 \%$ underestimated uncertainty remains in DR3, as do small ($\lesssim 0.03\,\rm mas\,yr^{-1}$) local frame rotations that depend on source position, color, and magnitude. 

After calibration of the reference frames, Brandt's analysis yields three proper motion measurements for each source observed by {\it Gaia} and {\it Hipparcos}: one at epoch $\approx 1991.25$, as measured by {\it Hipparcos}, one at epoch $\approx 2003.625$ (but averaged over $\sim 25$ years), calculated from the difference in the {\it Hipparcos} and {\it Gaia} positions, and one at epoch $\approx 2016$, calculated from {\it Gaia} data alone. These three proper motions essentially yield two acceleration measurements. \citet{Brandt2019} and \citet{Rickman2022} showed that for binaries with periods comparable to the {\it Hipparcos}-{\it Gaia} baseline, these acceleration measurements can yield useful constraints on companion masses when combined with a few epochs of relative astrometry, even when the latter only sample a small fraction of the orbit. 

An important caveat to astrometric companion mass constraints from PMa and/or \texttt{ruwe} is that both methods assume deviations from linear proper motion are purely gravitational, and are most easily interpreted when the companion is dark or at least much fainter than the luminous star. Companions that are luminous complicate analysis in two ways: (a) the measured acceleration becomes the acceleration of the photocenter rather than of the primary, which in general leads to underestimated companion masses, and (b), marginally resolved pairs of luminous stars will in general have poor proper motion measurements and underestimated uncertainties. This means that many binaries with separations of 0.2-2 arcsec will have large PMa and \texttt{ruwe}, even when the physical separations are too large for acceleration to be detectable. Such sources can in many cases be identified by \texttt{ipd\_frac\_multi\_peak} $>$ 0 and/or follow-up with high-resolution imaging.

\section{Orbital solutions in Gaia DR3}
\label{sec:orbits}
\begin{figure*}[!ht]
    \centering
    \includegraphics[width=\textwidth]{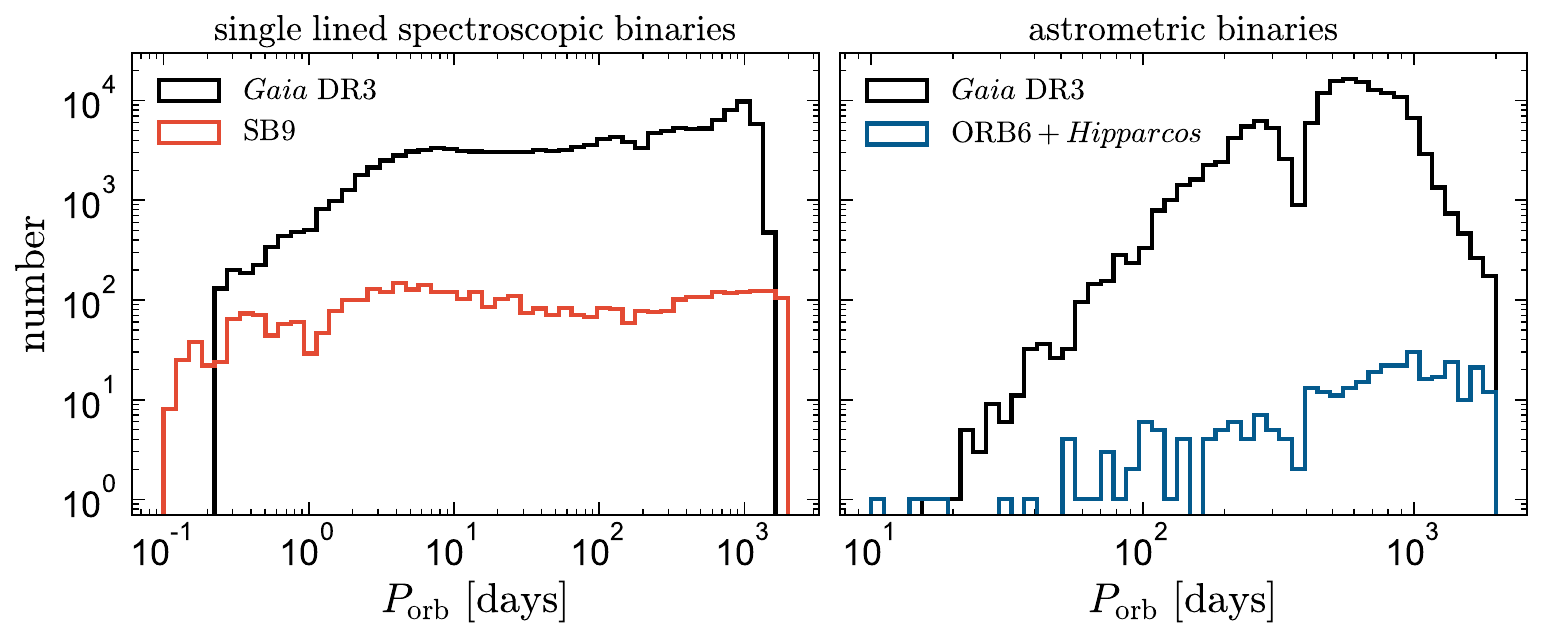}
    \caption{Orbital period distributions of {\it Gaia} DR3 single-lined spectroscopic binaries (left) and astrometric binaries (right). I compare to the largest sample of spectroscopic binaries before {\it Gaia} \citep[SB9, with $N \approx 5,000$;][]{Pourbaix2004, Merle2023} and to the largest pre-{\it Gaia} sample of astrometric binary orbits (ORB6 and {\it Hipparcos}, with $N \approx 300$ at $P_{\rm orb} < 1000$ days; \citealt{Hartkopf2001}). For binaries with full orbital solutions, {\it Gaia} represents a factor of $\sim 50$ increase in sample size {\it over all previous work}. }
    \label{fig:gaia_sample_size}
\end{figure*}

{\it Gaia} DR3 was the first data release to include orbital solutions for binaries. The quality cuts imposed in DR3 were conservative, and the published solutions represent only a small fraction ($\lesssim 1\%$; \citealt{Soderhjelm2004}) of those that should achievable by the end of the mission. Nevertheless, both the published spectroscopic and astrometric solutions represent more than an order of magnitude increase in sample size over the previous literature (Figure~\ref{fig:gaia_sample_size}). The published astrometric, spectroscopic, and light curve solutions are summarized in \citet{GaiaCollaboration2023_teaser}.

\subsection{Astrometric binaries}
\label{sec:ast}

\begin{figure*}[!ht]
    \centering
    \includegraphics[width=\textwidth]{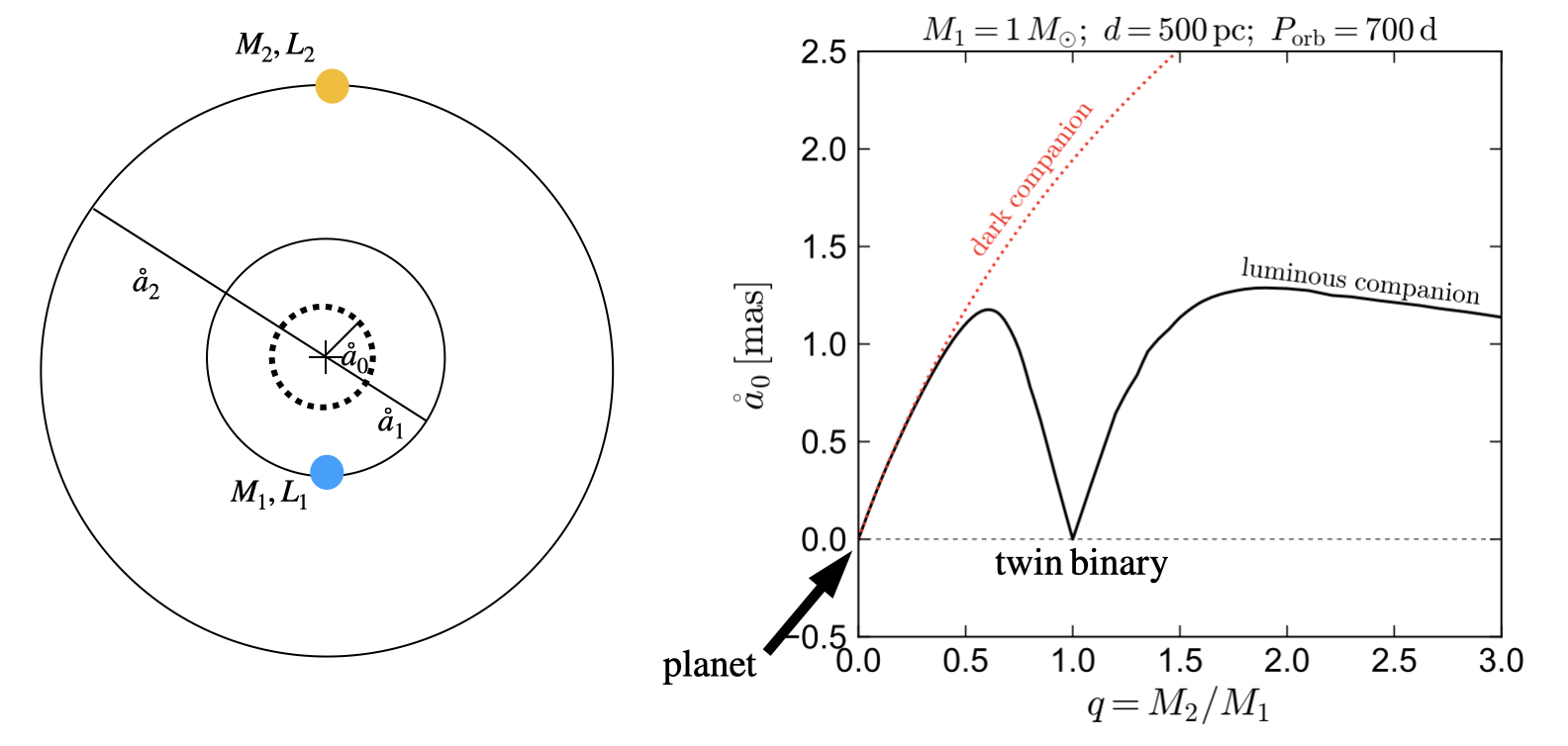}
    \caption{Left: {\it Gaia's} view of an astrometric binary. The system contains two stars, with masses $M_1$ and $M_2$, which in general can both contribute light. The angular separation between the stars is too small for {\it Gaia} to resolve, so what {\it Gaia} observes is the photocenter (dashed circle). Because the photocenter is between the two stars, the ellipse it traces out on the sky is in general smaller than the semi-major axis. Right: predicted photocenter semi-major axis for a binary containing a $1\,M_{\odot}$ main-sequence star at a distance of 500\,pc, with a 700\,d orbital period. Black line shows a range of main-sequence luminous companions, while red dashed line shows a dark companion. }
    \label{fig:astrometric}
\end{figure*}

Orbital solutions for astrometric binaries will perhaps be the most important and unique binary-related product of the {\it Gaia} mission. Astrometry has several advantages compared to other methods of detecting binaries: (a) it constrains the full orbital ellipse, including the inclination, (b) sensitivity to astrometric orbits is in principle easy to forward-model, leading to a better-understood selection function than spectroscopic orbits, and (c) astrometric wobble can be measured ``for free'' for all sources {\it Gaia} observes, including many sources that are too faint for {\it Gaia} to measure RV orbits. 

There are two primary tradeoffs. First, astrometry is sensitive to a relatively narrow range of orbital periods: $\sim 100-1000$ days in DR3. Second, sensitivity is poor at large distances. This makes astrometric orbits sub-optimal for studying rare objects such as massive stars, which are bright but mostly found at distances of several kpc. The sensitivity of astrometric orbits is similar to that of \texttt{ruwe} -- orbits can only be fit if deviations from a single-star model are significant -- but the requirements on phase coverage for a well-constrained solution are more restrictive.

The amplitude of the astrometric wobble depends in non-trivial ways on the mass and light ratio of a binary, as illustrated in Figure~\ref{fig:astrometric}. {\it Gaia} does not resolve the two stars, but instead sees the motion of their photocenter; i.e., the light-weighted average position in the $G-$band. The photocenter's motion as a function of time can be described by a Keplerian ellipse with angular scale $\mathring{a}_0$ \citep[see][for a historical and pedagogical review]{vandeKamp1975}.\footnote{Note that, due to projection effects, $\mathring{a}_0$ is {\it not} the semimajor axis of the photocenter orbit measured on the plane of the sky. Rather, $\mathring{a}_0$ is the physical 3D size of the orbit divided by distance, with appropriate scaling by the light ratio. This distinction can be appreciated by considering the case of a highly eccentric orbit viewed end-on, which can have a large $\mathring{a}_0$ while producing a small projected ellipse on the sky. }

Let us first consider a star with a dark companion, such as a planet or a compact object. In this case, the photocenter simply traces the star. If the binary orbit has semimajor axis $a$, the semimajor axis of the star's orbit is  $a_{1}=\frac{q}{1+q}a$, where $q=M_2/M_1$ and $M_1$ and $M_2$ are the masses of the brighter and fainter components. Then the ratio of the star's angular orbit, $\mathring{a}_1$, to the parallax is 
\begin{equation}
    \label{eq:ratio_a_parallax}
    \frac{\mathring{a}_{1}}{\varpi}=\frac{a_{1}}{1\,{\rm au}},
\end{equation}
which is independent of distance because the distance dependencies in $\mathring{a}_1$ and $\varpi$ cancel out. For a dark companion, $\mathring{a}_0 = \mathring{a}_1$.
Then in the limit of $q\ll 1$ (e.g., a planet), $\mathring{a}_{0}=q\varpi a$, while in the limit of $q\gg 1$ (e.g., a black hole) $\mathring{a}_{0}=\varpi a$. 

In the case of a luminous companion, the scale of the photocenter orbit depends on both the mass and light ratio: $\mathring{a}_{0}=\varpi\delta_{q\ell}a$, where 
\begin{equation}
    \label{eq:deltaql}
    \delta_{q\ell}=\frac{\left|q-\ell\right|}{\left(q+\ell\right)\left(\ell+1\right)}.
\end{equation}
Here $\ell=F_2/F_1$ is the $G-$band secondary-to-primary flux ratio. The right panel of Figure~\ref{fig:astrometric} shows how $\delta_{q\ell}$ depends on the mass ratio for typical main-sequence stars: using a 300-Myr old MIST isochrone with solar metallicity \citep{Choi2016}, I plot the predicted $\mathring{a}_0$ for a Sun-like star with a range of main-sequence companions. At mass ratios $q \lesssim 0.4$, the secondary contributes so little light that the photocenter essentially tracks the primary ($\ell \approx 0$). At higher $q$, the photocenter orbit becomes much smaller than the true orbit, and at $q=1$ (a twin binary), $\delta_{q\ell} \to 0$. At $q > 1$, the light becomes dominated by the other star, whose wobble due to the Sun-like star becomes progressively smaller as $q$ increases. 

{\it Gaia} astrometric orbits are parameterized using the Thiele-Innes angles \citep[$A$, $B$, $F$, and $G$; e.g.][]{vandeKamp1975}. These quantities, which are transformations of the more familiar orbital angles $i$, $\mathring{a}_0$, $\Omega$, and $\omega$, are advantageous for orbit fitting because their optimization (and that of the standard 5 astrometric parameters) is linear once the period, eccentricity, and periastron time are specified. Astrometric binary solutions thus have 12 free parameters, only 3 of which require nonlinear optimization \citep{Pourbaix2022}.

\begin{figure*}[!ht]
    \centering
    \includegraphics[width=\textwidth]{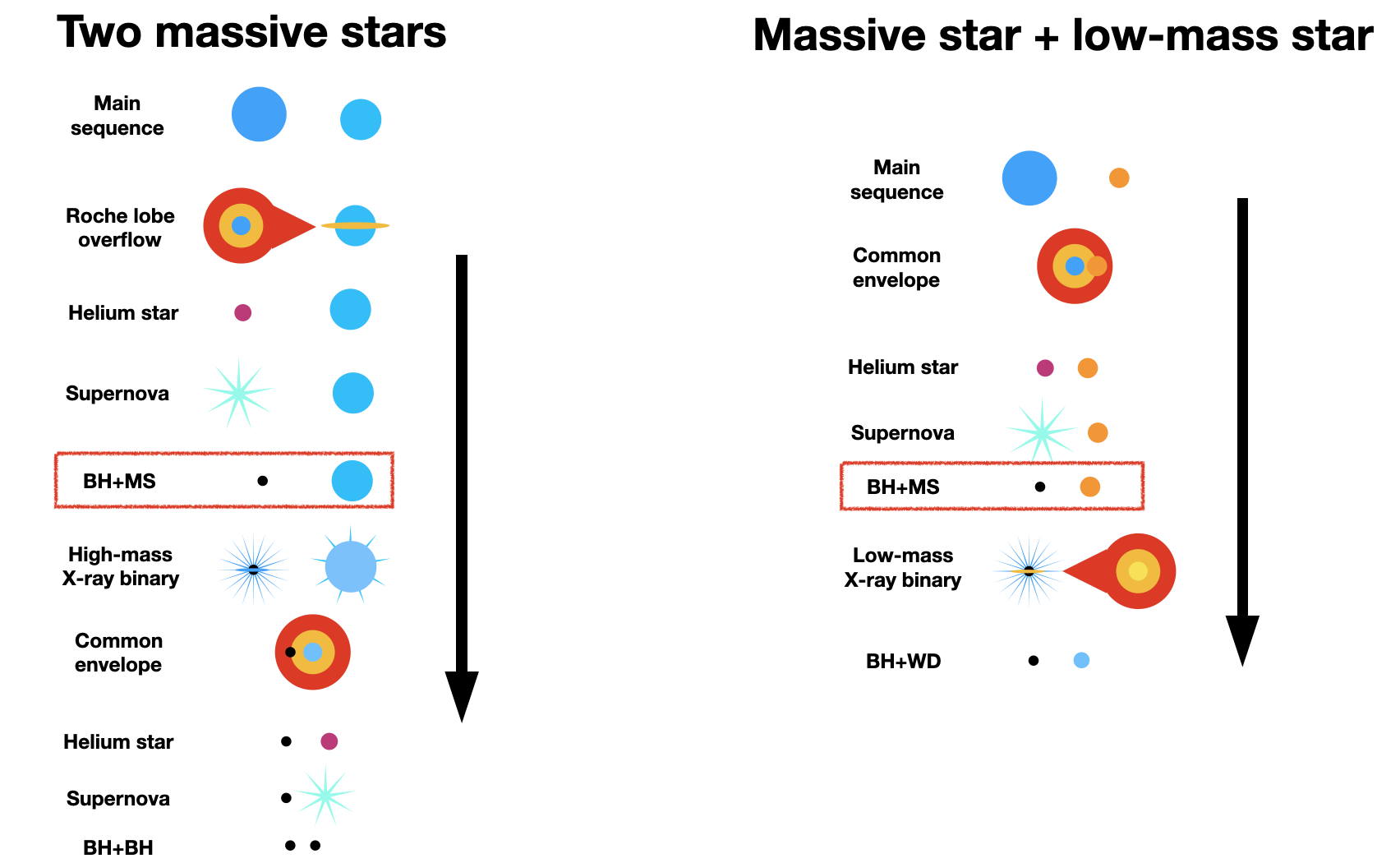}
    \caption{Schematic evolutionary models for massive stars with high- and low-mass  companions. Both models include a phase in which a non-accreting BH is orbited by a main-sequence star (red boxes). Some such systems eventually evolve to become high-mass X-ray binaries and BH+BH mergers (for two massive stars) and low-mass X-ray binaries (for massive stars with low-mass companions). {\it Gaia} can discover non-interacting binaries in the BH+MS phase with both high- and low-mass companions.}
    \label{fig:BH_evol}
\end{figure*}

It follows from Equation~\ref{eq:ratio_a_parallax} that for dark companions, astrometric wobble will be larger than parallax wobble if $a_1 > 1\,{\rm au}$. Given the $\sim$1000-day observing baseline of DR3, this implies that some binaries -- particularly those with dark and massive companions -- will have their orbits constrained with higher relative precision than their parallaxes. In future data releases, we can also expect to detect binaries that have well-measured orbits even with insignificant parallaxes. No such solutions were included in DR3, primarily as a result of conservative quality cuts that were employed \citep{Halbwachs2023}. The strictest cut -- which leads to the steep decline in the sensitivity of astrometric orbits at short periods shown in Figure~\ref{fig:detection_methods} -- was:

\begin{equation}
    \label{eq:par_over_err}
    \frac{\varpi}{\sigma_\varpi} > \frac{20,000\,\rm d}{P_{\rm orb}}.
\end{equation}

As a result, almost all solutions published in DR3 have $\varpi/\sigma_{\varpi} \gtrsim 20$, and, for example, solutions with $P_{\rm orb} \sim 100$ d are only included if they have  $\varpi/\sigma_{\varpi} > 200$. This rather harsh filtering was employed in DR3 to eliminate what would otherwise have been serious contamination with spurious solutions. Astrometric binary solutions were also not calculated for sources with \texttt{ipd\_frac\_multi\_peak > 2}, indicative of nearby resolved or marginally resolved companions; this disprivileges triples. Spurious solutions will continue to exist in DR4 and DR5, but the epoch-level astrometric data will make it possible to identify and vet lower-significance solutions that are of astrophysical interest.  

DR3 included astrometric orbital solutions for 168,065 sources, including 134,598 with purely astrometric solutions (\texttt{nss\_solution\_type = Orbital}) and 33,467 joint astrometric + RV solutions (\texttt{nss\_solution\_type = AstroSpectroSB1}). In addition, astrometric solutions including acceleration were published for 338,215 sources. Many of these are likely binaries with periods $\gtrsim 1000$ d, while some are shorter-period binaries for which the astrometric pipeline failed to find the correct solution \citep{Pourbaix2022}. A small number of sources were processed with other astrometric models, as described by \citet{GaiaCollaboration2023_teaser} and \citet{Holl2023}. In no case was the epoch astrometry published. Some spurious solutions can be identified based on their quality flags (e.g., many but not all solutions with \texttt{goodness\_of\_fit} $\gtrsim 5$ appear to be spurious) and periods close to values related to the {\it Gaia} scanning law \citep{Holl2023b}.

The acceleration solutions have not yet seen much use in the literature. In principle, these solutions contain similar information to PMa (Section~\ref{sec:prop_motion_anomaly}) and are simpler to interpret because there is no need to calibrate the reference frames of different missions. Acceleration solutions may eventually become powerful for identifying stellar- or even intermediate-mass black holes from partial orbits \citep{Andrews2023}.

\subsubsection{The hunt for black holes}
\label{sec:compact_objects}

There has been great interest over the last decade in the population of stellar-mass BHs that {\it Gaia} will discover in astrometric binaries. Such systems offer an opportunity to study a BH population selected in a different way from BHs in X-ray binaries and gravitational wave events. {\it Gaia}-detected BH binaries may nevertheless evolve to become X-ray binaries and gravitational wave sources (Figure~\ref{fig:BH_evol}), and so studies of the {\it Gaia}-detected systems can constrain models for the formation of X-ray binaries and gravitational wave sources.

 Under single-star evolution, most BH progenitors are expected to reach radii of several au as red supergiants, such that they cannot fit inside orbits with $a\lesssim 10$ au without overflowing their Roche lobes, leading to stable mass transfer or common envelope evolution. This suggests that most BHs in binaries detectable by {\it Gaia} will be products of binary mass transfer, though some systems could avoid interaction if they formed through dynamical interactions in dense clusters, or in wider binaries. The outcome of binary interactions and the BH populations of clusters are both uncertain, and so the BH population {\it Gaia} is expected to discover is also uncertain. 

Early predictions for {\it Gaia's} yield of compact object binaries were optimistic. For example, \citet{Mashian2017} predicted that {\it Gaia} will discover $2\times 10^5$ BHs in binaries with orbital periods shorter than 5 years. They neglected binary evolution, assuming that BH + luminous star binaries have a log-uniform separation distribution and that most stars with initial masses above $20\,M_{\odot}$ form BHs with companions that survive the massive star's evolution and death.

\begin{figure*}[!ht]
    \centering
    \includegraphics[width=\textwidth]{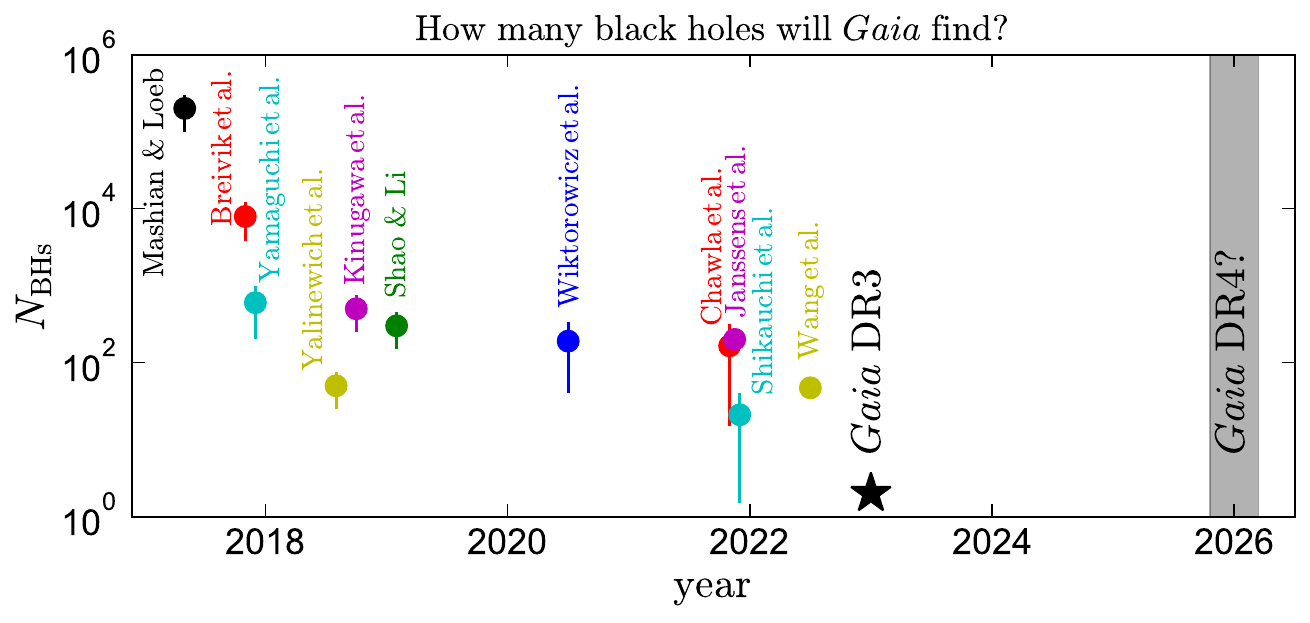}
    \caption{Predictions from several population synthesis calculations (see Section~\ref{sec:compact_objects}) for the number of black holes {\it Gaia} will find. The actual number discovered in DR3, $N_{\rm BHs} = 2$, is shown with a black star. Predictions have generally gotten more pessimistic over time but were all optimistic compared to the actual number discovered so far. However,  only orbital solutions with high astrometric signal-to-noise were published in DR3, so DR4 (planned for early 2026) is expected to enable the discovery of a significantly larger sample.}
    \label{fig:NBHs}
\end{figure*}

\citet{Breivik2017} made the first attempt to incorporate binary evolution via population synthesis simulations in predictions for the {\it Gaia} BH population. They predicted $\sim 3,800-12,000$ BHs to be detectable with {\it Gaia}, reflecting the fact that simulations predict a large fraction of would-be BH binaries to be disrupted during or before the formation of the BH. Despite this reduction, their predicted BH population would represent a 100-fold increase in the number of known stellar-mass BHs. An update on these population synthesis calculations was presented by \citet{Chawla2022}, who revised the predicted number of detectable BHs downward by more than an order of magnitude, to $\sim 30-300$ systems. These revisions owed to accounting for dust extinction, changes in the adopted metallicity-dependent star formation history of the Milky Way, and various changes in the binary evolution modeling.

Two BHs were discovered among the orbital solutions published in {\it Gaia} DR3. The first, Gaia BH1, contains a solar-type star in a 185-day orbit around a $9.3\,M_\odot$ dark companion \citep{El-Badry2023_bh1, Chakrabarti2023}. The combination of precise radial velocity measurements and {\it Gaia} astrometry allows the 3D orbit and companion mass to be robustly constrained. The system's RV mass function is high enough that any plausible companion besides a BH can be ruled out even on the basis of RVs alone. The 2nd system, Gaia BH2, also contains a $\sim 1\,M_{\odot}$ star and a $\sim 9\,M_{\odot}$ BH \citep{Tanikawa2023, El-Badry2023_bh2}. The orbital period is 1277 days, about 30\% longer than the {\it Gaia} DR3 observational baseline. The star is somewhat evolved ($R\approx 8\,R_{\odot}$) and thus more luminous than a MS star. This allowed multi-epoch RVs to be measured by {\it Gaia}, and its \texttt{AstroSpectroSB1} solution is based on both RVs and astrometry.

A third system, Gaia BH3, was discovered and published in the course of processing pre-release Gaia DR4 astrometry \citep{GaiaCollaboration2024}. This system contains a very low metallicity star ($[\rm M/H]\approx -2.2$) on a halo orbit, and a $33\,M_{\odot}$ BH. Gaia BH3 contains the most massive robustly-measured stellar-mass BH by a significant margin. It provides the first empirical evidence that low-metallicity massive stars leave behind massive BHs, as has long been assumed in modeling the progenitors of gravitational wave sources. The systems has an 11.6-year orbital period -- too long to have been solved with DR3 astrometry -- and is only 600 pc from Earth. 

Many other population synthesis studies have made predictions for the BH population {\it Gaia} will discover \citep[e.g.][]{Yamaguchi2018, Yalinewich2018, Kinugawa2018, Shao2019, Wiktorowicz2020, Chawla2022, Janssens2022, Shikauchi2022, Wang2022}; some predictions are compared and summarized in Figure~\ref{fig:NBHs}. All these studies assumed formation through isolated binary evolution.

\begin{figure*}[!ht]
    \centering
    \includegraphics[width=\textwidth]{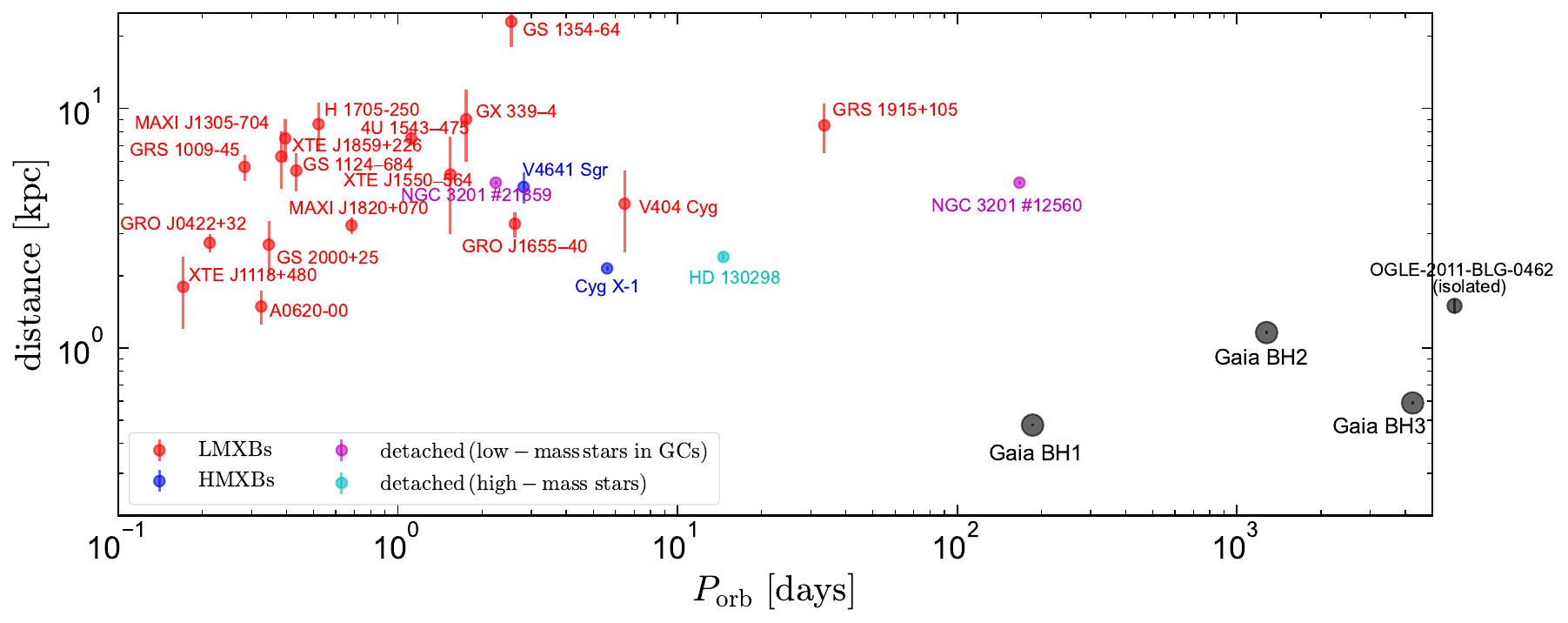}
    \caption{Comparison of dormant BHs discovered with {\it Gaia} (black) to other known BHs. Red and blue symbols correspond to accreting BHs with low- and high-mass companions. Magenta symbols show detached binaries in the globular cluster NGC 3201, and cyan points show detached binaries in which the luminous star is a high-mass ($\gtrsim 20\,M_{\odot}$) star. The {\it Gaia}-discovered BHs stand out from the rest of the population due to their long periods. All three are closer to Earth than any other systems in the plot, implying that such wide BH binaries significantly outnumber X-ray binaries. The single isolated BH detected via astrometric microlensing is shown outside the axis limits.    }
    \label{fig:bh_distance}
\end{figure*}

Predictions have overall become more pessimistic over the last decade, but several studies shortly before {\it Gaia} DR3 predicted a few hundred of BHs -- significantly more than the $N_{\rm BHs} = 2$ detected in DR3 (or $N_{\rm BHs} = 3$ if Gaia BH3 is included).  This likely owes in large part to a mismatch between the detectability thresholds assumed in population synthesis studies and the cuts actually employed for solutions published in DR3. The situation will improve in {\it Gaia} DR4, when epoch-level astrometry for all sources will be published and the community will be able to choose its own quality cuts. I consider the detectability thresholds assumed by most of the studies in Figure~\ref{fig:NBHs} to be relatively optimistic. For example, \citet{Chawla2022} adopt an ``optimistic'' detection threshold of $\mathring{a}_{1} > \sigma_\xi$, and a ``pessimistic'' threshold of $\mathring{a}_{1} > 3\sigma_\xi$, where $\sigma_\xi$ is the typical single-epoch precision of {\it Gaia} observations in the along-scan direction at a given apparent magnitude.  

For Gaia BH1, BH2, and BH3, $\mathring{a}_{1}\approx 23 \sigma_\xi$, $\mathring{a}_{1}\approx 53 \sigma_\xi$, and $\mathring{a}_{1}\approx 500 \sigma_\xi$, respectively. The BHs discovered so far thus have astrometric orbits that are constrained $\sim 20-500$ times better (before RV follow-up) than \citet{Chawla2022} assume orbits need to be constrained in order to identify binaries as containing BHs. This implies that either DR4 will yield of order $100$ more systems like Gaia BH1 and BH2, or higher-SNR astrometry is required to effectively distinguish BH candidates from the noise. Reality likely falls between these two scenarios.

The Gaia BHs are now the three closest known BHs (Figure~\ref{fig:bh_distance}). The Copernican principle thus suggests that wide, non-accreting BHs are significantly more common than their X-ray-bright cousins \citep{Rodriguez2023}.
How these systems formed is, however, still uncertain: binary evolution models predict that the red supergiant + low-mass star binaries would only survive common envelope evolution if the low-mass star spirals in to a short-period orbit, similar to observed BH low-mass X-ray binaries. Several wide neutron star binary candidates have also been discovered \citep{El-Badry2024_ns1, El-Badry2024_ns_sample}, whose formation is difficult to explain for similar reasons.

Triple evolution offers some scenarios to avoid a common envelope \citep{Generozov2023}, but triple models predict a larger BH + MS binary population at shorter periods, which would be detectable with {\it Gaia} RV orbits (Section~\ref{sec:rv_orbits}) and has not yet been observed. High-precision RVs for Gaia BH1 disfavor a scenario in which the system is a triple today \citep{Nagarajan2024}. Another possibility is that the binaries formed through dynamical exchange in dense clusters \citep{Rastello2023, DiCarlo2023, Tanikawa2024, MarinPina2024}. This scenario seems particularly promising for Gaia BH3, with is part of the ED-2 stellar stream \citep{Dodd2023, Balbinot2023, Balbinot2024}.

Isolated BHs, which can currently only be studied with microlensing, likely outnumber wide BH binaries. However, these systems can only be studied when they fortuitously pass in front of a background star, and masses can only be measured with precision astrometry. To date, only one high-confidence BH has been discovered in this way \citep[Figure~\ref{fig:bh_distance};][]{Sahu2022, Lam2022, Mroz2022, Lam2023}; epoch astrometry from future {\it Gaia} releases may reveal a few more \citep[e.g.][]{Belokurov2002, Rybicki2018, Kruszynska2024, Howil2024}.

\begin{figure*}[!ht]
    \centering
    \includegraphics[width=\textwidth]{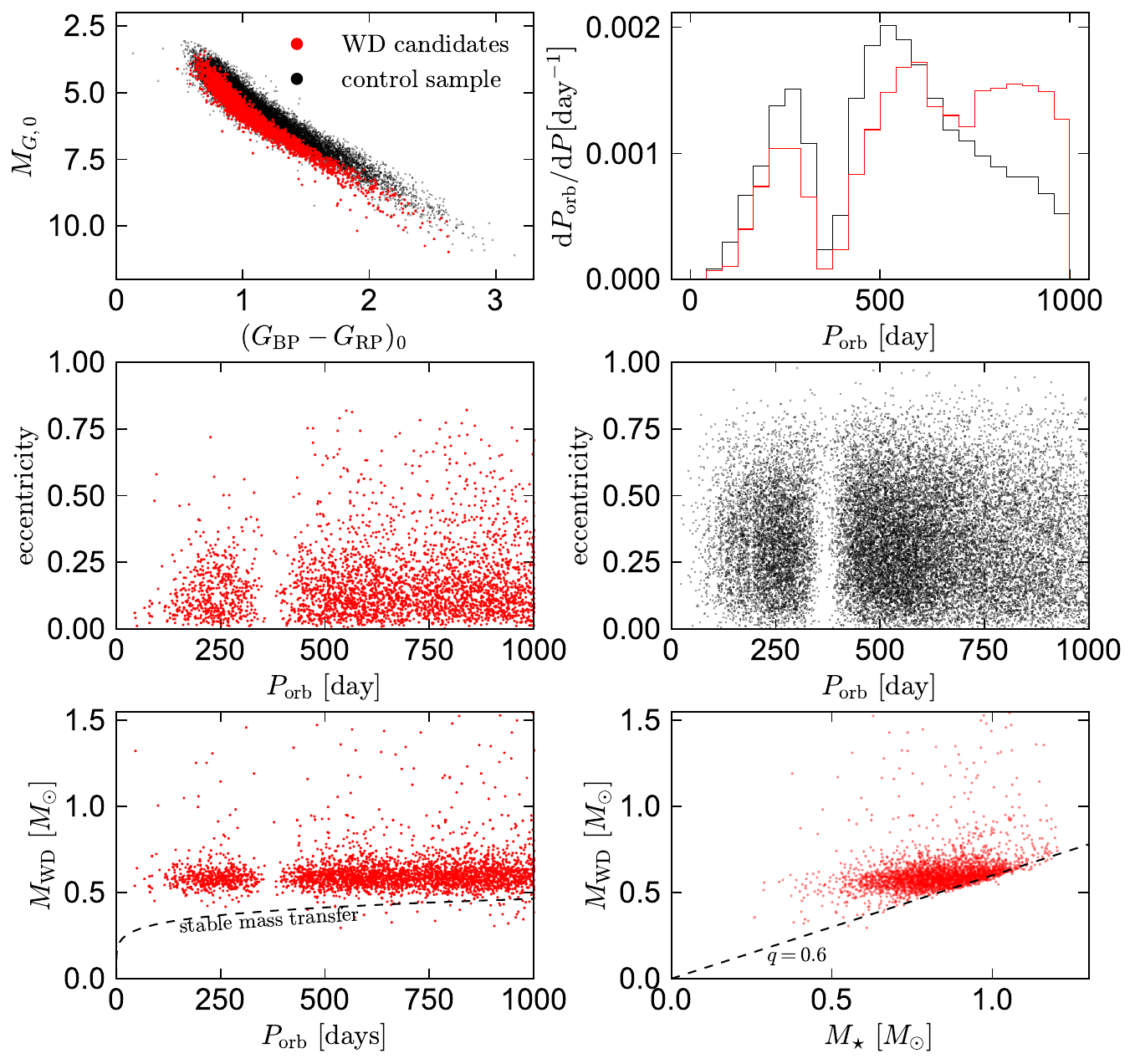}
    \caption{The population of white dwarf + main sequence binary candidates identified by \citet[][red]{Shahaf2023b}, compared to a control sample of astrometric binaries selected to have the same distance and absolute magnitude distribution (black). The median distance is 470 pc. Compared to the control sample, the WD binaries have low eccentricities (middle row), as expected if their orbits were tidally circularized when the white dwarfs' progenitors were red giants. The period distributions of the control sample and the WD binaries are remarkably similar, mostly reflecting the {\it Gaia} selection function, but the WD binaries display a modest dearth of short periods ($P_{\rm orb} \lesssim 1$\,yr) and excess of long periods. The sample is deficient in low-mass WDs for $M_\star \lesssim 0.8\,M_{\odot}$ due to the astrometric selection procedure (lower right). The WD mass distribution is not strongly correlated with period and is inconsistent with the $P_{\rm orb}-M_{\rm WD}$ relation expected for binaries formed by stable mass transfer \citep[lower left;][]{Rappaport1995}.}
    \label{fig:wds}
\end{figure*}

\subsubsection{White dwarfs}
In addition to black holes and neutron stars, {\it Gaia} DR3 yielded a much larger sample of WDs in astrometric binaries.  Using the algorithm originally laid out in \citet{Shahaf2019}, \citet{Shahaf2023} selected a sample of $\sim 150$ WD+MS binary candidates in which the astrometrically-implied minimum mass ratio was too large for the companion to be a MS star {\it or} an unresolved binary containing two MS stars. Their sample was dominated by massive WDs, because the hierarchical triple scenario can only be ruled out when the mass ratio is $M_2/M_1 \gtrsim 1$. Subsequently, \citet{Shahaf2023b} showed that a much larger sample of WDs can be selected if the astrometric orbit and position in the CMD are considered together. While a companion that is itself a tight binary can produce a large astrometric mass function, it will also contribute light, making the unresolved source overluminous compared to a MS star of the same color. \citet{Shahaf2023b} thus selected all $\sim 10,000$ astrometric binaries with photocenter wobbles too large to be explained by a single MS companion, and then rejected the $\sim 7,000$ in which the unresolved sources is above the main sequence. The result is a catalog of $\sim 3,000$ binaries mostly containing WD companions. Low-mass WDs are still under-represented overall (these can best be characterized after having been identified via UV excess; e.g. \citealt{Garbutt2024}), but this approach allows them to be identified as WDs with mass ratios down to $M_{\rm WD}/M_\star \gtrsim 0.6$.  

Properties of the WD+MS astrometric binary sample from \citet{Shahaf2023b} are explored in Figure~\ref{fig:wds}, where I compare their distributions to those of a control sample of astrometric binaries selected to have the same distance and absolute magnitude distributions. The control sample presumably consists mostly of MS+MS binaries. The fact that eccentric orbits are underrepresented in the WD binary sample compared to the control sample validates that most of these objects are really WDs: tides are expected to circularize the orbits of WD binaries when the WD progenitors are red giants. Indeed, it is still a mystery why most of these binaries have low but nonzero eccentricities. 

The period distribution of both WD binaries and the control sample displays a dip at $P_{\rm orb}\sim 1$\,yr and reduced sensitivity at short periods (due to a smaller astrometric ellipses at fixed distance) and long period (due to the finite duration of the {\it Gaia} observing baseline). When these effects are accounted for, the period distribution of the WD + MS binaries revealed by {\it Gaia} is surprisingly similar to that of the control sample: the combined effects of common envelope evolution and/or stable mass transfer modify the period distribution (in the separation range {\it Gaia} is sensitive to) in undramatic ways. The mass distribution of the WDs is rather sharply peaked at $0.6\,M_{\odot}$, with fewer high-mass WDs found in a volume-limited sample than in the field \citep{Hallakoun2023}. This may be a result of a dearth of merger products in the astrometric binary sample, or a result of the mass transfer process. The period--WD mass relation for most of these WDs is manifestly inconsistent with predictions for stable mass transfer (Figure~\ref{fig:wds}, lower left). This is partially a selection effect, since most WDs formed by stable mass transfer have too low masses to be identified as WDs based on astrometry alone \citep[e.g.][]{Garbutt2024}, but the question of how these WDs ended up in the orbits where they are observed remains.

Most of the WD companions identified in astrometric orbits have separations of order 1-3\,au. This is in some sense not surprising, because this is the separation range to which {\it Gaia} DR3 was most sensitive. It is surprising, however, in the sense that only a few WD companions in comparable orbits were known previously \citep[e.g.][]{Kawahara2018, Masuda2019, Yamaguchi2024b}. Moreover, simple population synthesis simulations predict that the region of the period--mass parameter space occupied by the {\it Gaia} WD binaries should be barren, having been cleared out by common envelope evolution when the progenitors of the WDs were red giants \citep[see][]{Shahaf2023b}. A plausible explanation is that these systems are the result of unstable mass transfer from AGB star donors \citep{Yamaguchi2024, Belloni2024}.

\subsubsection{Planets and brown dwarfs}
\label{sec:planets}
{\it Gaia} has been predicted to eventually discover between $10^4$ and $10^5$ exoplanets via astrometric wobble of their host stars \citep{Perryman2014}. The largest astrometric signal is produced by high-mass planets orbiting low-mass stars in periods comparable to the {\it Gaia} observing baseline. Optimistic assumptions about SNR thresholds for a confident detection may be more reasonable for low-mass companions than for BH companions, simply because planets and brown dwarfs are common. A modest number of false positives are thus unlikely to dominate the observed signal, as they would BH companion candidates that are not carefully vetted. 

Astrometric solutions for a few dozen high-mass planets or low-mass brown dwarfs were already published in {\it Gaia} DR3. With astrometry alone, it is hard to distinguish such low-mass companions from nearly equal-mass twin binaries (Figure~\ref{fig:astrometric}). However, twins are overluminous by a factor of two compared to the main sequence and are generally double-lined, so they can be filtered out with modest follow-up \citep[e.g.][]{Marcussen2023}. 

\citet{Winn2022} showed that {\it Gaia} astrometric solutions for planets can be combined with ground-based RVs, yielding tighter constraints on planet mass than can be achieved with either dataset alone. \citet{Marcussen2023} used a similar approach and reported success in some cases, but also found cases in which the true RV semi-amplitude was significantly smaller than predicted by the astrometric solution. \citet{Winterhalder2024} showed that the widest astrometric binaries, with separations $\gtrsim 30\,{\rm mas}$, can be resolved with GRAVITY interferometry. Like RVs, interferometric follow-up can dramatically tighten {\it Gaia} constraints on orbits and component masses. 

Joint fitting of astrometry and RVs for previously known candidates has shown many candidate giant planets to actually be brown dwarfs, and many candidate brown dwarfs to actually be low-mass stars \citep{Unger2023, Fitzmaurice2023, Stevenson2023, Xiao2023}. {\it Gaia} DR3 data reaffirms the existence of a brown dwarf desert, which appears to be most barren at masses of $35-55\,M_{\rm Jup}$ and periods of $\lesssim 10^3$ days, where the completeness is relatively well-understood.

\subsection{Spectroscopic binaries}
\label{sec:rv_orbits}
{\it Gaia} DR3 included orbital solutions based purely on RVs for 181529 single-lined binaries (\texttt{nss\_solution\_type}=  \texttt{SB1} or \texttt{SB1C}) and 5376 double-lined binaries (\texttt{nss\_solution\_type}= \texttt{SB2} or \texttt{SB2C}). In addition, 56808 ``trend'' solutions were fit for sources whose RVs showed evidence for acceleration but were not sufficient for determination of an unambiguous orbital solution. These solutions were designed to fit binaries with periods longer than the $\sim 1000$ day baseline of observations included in DR3. 

Some properties of the RV solutions are described by \citet{GaiaCollaboration2023_teaser}, but the processing of the epoch RV data and fitting of the orbits has not yet been described in a publication. Epoch RVs for a small fraction of these sources -- primarily those that were processed by the long period variable pipeline but are actually giants displaying ellipsoidal variability -- were published as part of the {\it Gaia} focused product release \citep{GaiaCollaboration2023_lpv, 2024arXiv240109531R}. An exploration of the RV orbits was also carried out by \citet{Merle2023}, who concluded that a large number of good SB1/SB2 orbits were filtered out because the goodness of fit metrics applied were overly conservative.  \citet{Bashi2022} showed that the rate of spurious orbital solutions among the spectroscopic orbital solutions is generally highest at short periods; they designed a quality metric to filter out spurious solutions that combines information contained in several different {\it Gaia} quality flags.

Information about the RV variability of objects with $G_{\rm RVS} < 12$ is published in DR3, even for cases where no orbital solution was calculated. The most useful flag is \texttt{rv\_amplitude\_robust}, which reports the peak-to-valley range of all the epoch RVs obtained for a source, after excluding outliers. Of course, some RV scatter is expected even for sources that are not binaries due to measurement uncertainties, so binaries are best identified as sources with  \texttt{rv\_amplitude\_robust} values significantly higher than typical sources of the same color and apparent magnitude. Rapidly rotating stars are a failure point for this type of analysis.

Spectroscopic binary orbits can only be fit for sources bright enough that RVs can be measured in single-epoch spectra. In DR3, solutions were attempted only for sources with $G_{\rm RVS} < 12$, which corresponds to a distance limit of $\sim  500$ pc for solar-type stars (Figure~\ref{fig:detection_methods}). RVs could also only be measured for stars with strong lines within the {\it Gaia} RVS bandpass with spectral types well-matched to the library of templates used for fitting RVs. In practice, this meant slowly-rotating stars for which the best-fit template had $T_{\rm eff}$ between 3875 and 8125\,K \citep{GaiaCollaboration2023}.

Whether a binary is single- or double-lined depends, in almost all cases, on the sensitivity and wavelength coverage of the observations. The {\it Gaia} solutions are based on RVs measure from  spectra that span a narrow wavelength range centered on the Ca II triplet. Lines of the secondary are unlikely to be detected in cases where it is faint or has a spectral type without strong lines in the RVS spectral bandpass. Validation of the {\it Gaia} spectroscopic orbits by comparison to other catalogs of spectroscopic binaries shows that many binaries recognized as double-lined in other catalogs received \texttt{SB1} orbits in {\it Gaia} DR3 \citep[e.g.][]{Tokovinin2023_spec, Merle2023}. The RV variability amplitudes for double-lined systems that were fit with SB1 solutions is often underestimated, probably as a result of blended lines.

\citet{Tokovinin2023} demonstrated that the presence of unresolved wide companions can lead to significant errors in RV orbital solutions as well as astrometric orbits, for multiple reasons. First, light from wide companions dilutes the inferred RV variability amplitude of spectroscopic binaries and the photocenter wobble of astrometric binaries, resulting in underestimated variability amplitudes. Second, the slitless, time-delay integrated nature of {\it Gaia} RVS spectroscopy requires the position of each source to be known accurately in order for the wavelength solution to be correct \citep{Boubert2019, Katz2023}. In unresolved and marginally resolved binaries, the source position in the focal plane can be different from what is assumed in processing the RVS spectra, leading to spurious velocity shifts on periods related to the scanning law.

%A search for unresolved wide companions is thus an important step in vetting orbital solutions of particular interest. 

\citet{Bashi2023} used a cleaned sample of spectroscopic orbits from DR3 to study the onset of tidal circularization in close binaries. They found, contrary to some previous work, that the circularization period depends weakly (if at all) on stellar age, and more strongly on effective temperature, with warmer stars being circularized at shorter periods. Of course, stellar ages are difficult to measure and correlated with effective temperature. 

Searches for compact object companions have also been carried out with the SB1 solutions, with several works investigating the nature of binaries with the highest RV mass functions \citep{ El-Badry2022_what_are, Fu2022, Jayasinghe2023, Rowan2024}. These searches have thus far revealed samples of high-mass white dwarfs \citep[e.g.][]{Yamaguchi2024} and many Algol-type mass-transfer binaries, but no high-confidence BHs or neutron stars. Contamination is generically higher in spectroscopically-selected samples of compact object candidates than in astrometrically-selected samples, because unrecognized light from a luminous secondary tends to increase rather than decrease the inferred secondary mass.

\subsection{Eclipsing and ellipsoidal binaries}
DR3 also included a sample of short-period binaries displaying photometric variability, including both eclipsing and ellipsoidal systems \citep{Mowlavi2023}. 86,918 sources have pure light curve solutions, while 155 have joint light curve + RV solutions. Larger samples of eclipsing binaries have been assembled elsewhere \citep[e.g.][]{Soszynski2016}, but the {\it Gaia} sample has the advantage of being all-sky. For most sources published in DR3, the $G-$band light curves have smaller per-epoch uncertainties than light curves produced by e.g. the ZTF and OGLE surveys, but have fewer photometric epochs. Light curves were classified with a machine learning classifier \citep{Rimoldini2023}.

As shown by \citet{Gomel2021}, a minimum mass ratio for an ellipsoidal variable can be derived from the variability amplitude, subject to the assumption that all observed variability is ellipsoidal. \citet{Gomel2023} used this fact to select a sample of $\sim 6000$ candidate ellipsoidal variables that may host a compact object companion, including 262 high-confidence systems with minimum mass ratios significantly larger than one. \citet{Nagarajan2023b} investigated the nature of this sample in more detail, carrying out spectroscopic follow-up on a sub-sample of the candidates. They showed that a majority of the sample are contact binaries containing two luminous stars, but a small fraction may still contain compact objects.

\begin{figure*}[!ht]
    \centering
    \includegraphics[width=\textwidth]{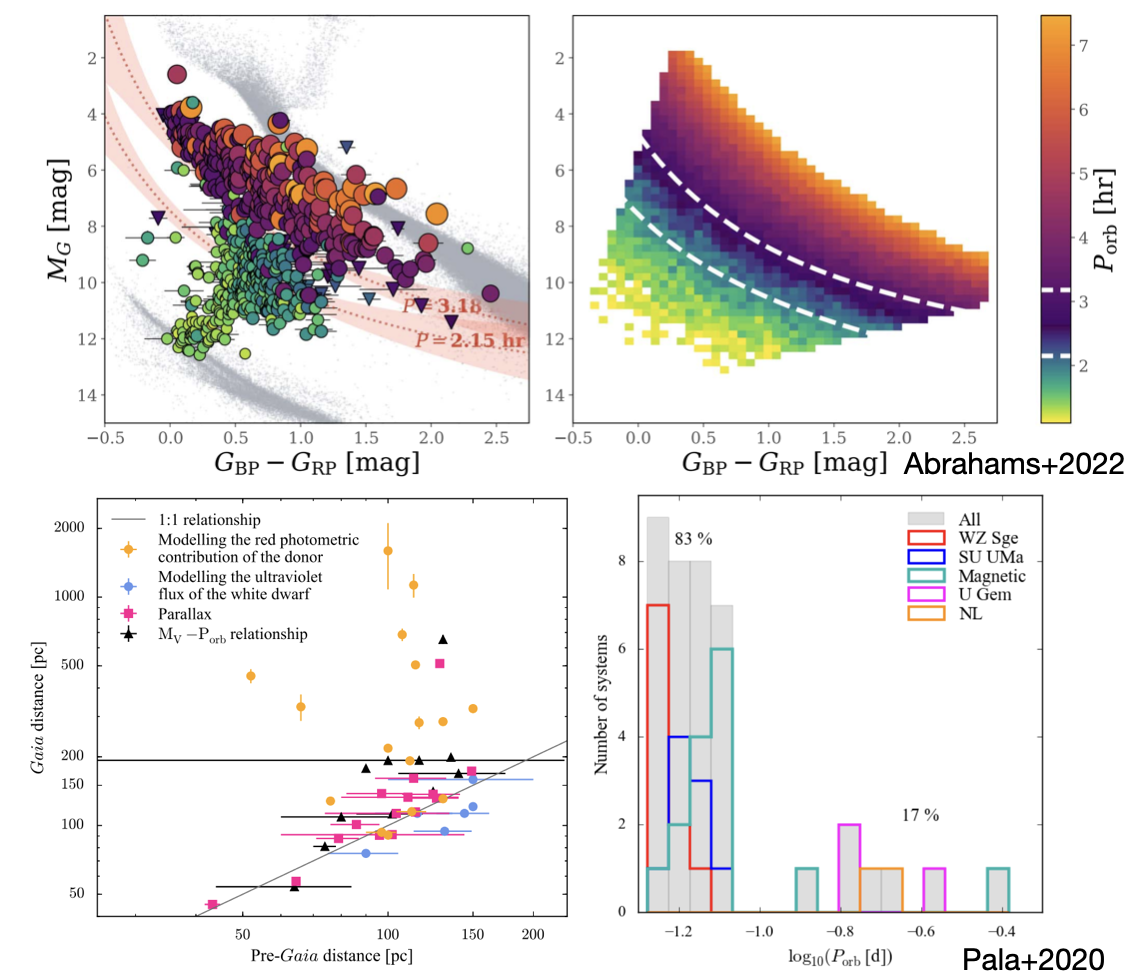}
    \caption{Top: cataclysmic variables on the {\it Gaia} color-magnitude diagram, adapted from \citet{Abrahams2022}. Colored circles show CVs with known orbital periods, compared on the CMD to a sample of nearby stars (gray). Points are colored and sized according to their period. The absolute magnitude of CVs decreases almost monotonically with orbital period, as is captured in the empirical model in the upper right panel.  Dashed white lines enclose the empirical CV period gap. Bottom: the 150 pc sample of CVs, adapted from \cite{Pala2020}. Left panel compares CV distances from {\it Gaia} DR2 parallaxes to previous estimates from the literature. Many CV distances were previously underestimated. Right panel shows the period distribution of CVs within 150 pc. It is dominated by sources below the period gap.}
    \label{fig:CVs}
\end{figure*}

\section{Distances for other binary samples and new discoveries}
\label{sec:cvs}
{\it Gaia} astrometry and RV measurements are ill-suited for characterizing compact binaries: they produce small astrometric wobble and have components whose RVs cannot be reliably measured from RVS spectra because they are too faint, rapidly rotating, or lack strong lines in the RVS bandpass. {\it Gaia} has nevertheless enabled significant progress in the study of these binaries by providing precise distance measurements. Exotic binary populations for which {\it Gaia} distances have improved understanding include  cataclysmic variables \citep{Pala2020, Abril2020, El-Badry2021_elms, Abrahams2022}, AM CVn binaries \citep{Ramsay2018, Kupfer2018, vanRoestel2022}, low- and high-mass X-ray binaries \citep{Gandhi2019}, ultracompact white dwarf binaries \citep{Burdge2020}, and binary millisecond pulsars \citep{Moran2023, Koljonen2023}. I discuss a few studies in more detail below. 

\begin{figure*}
    \centering
    \includegraphics[width=\textwidth]{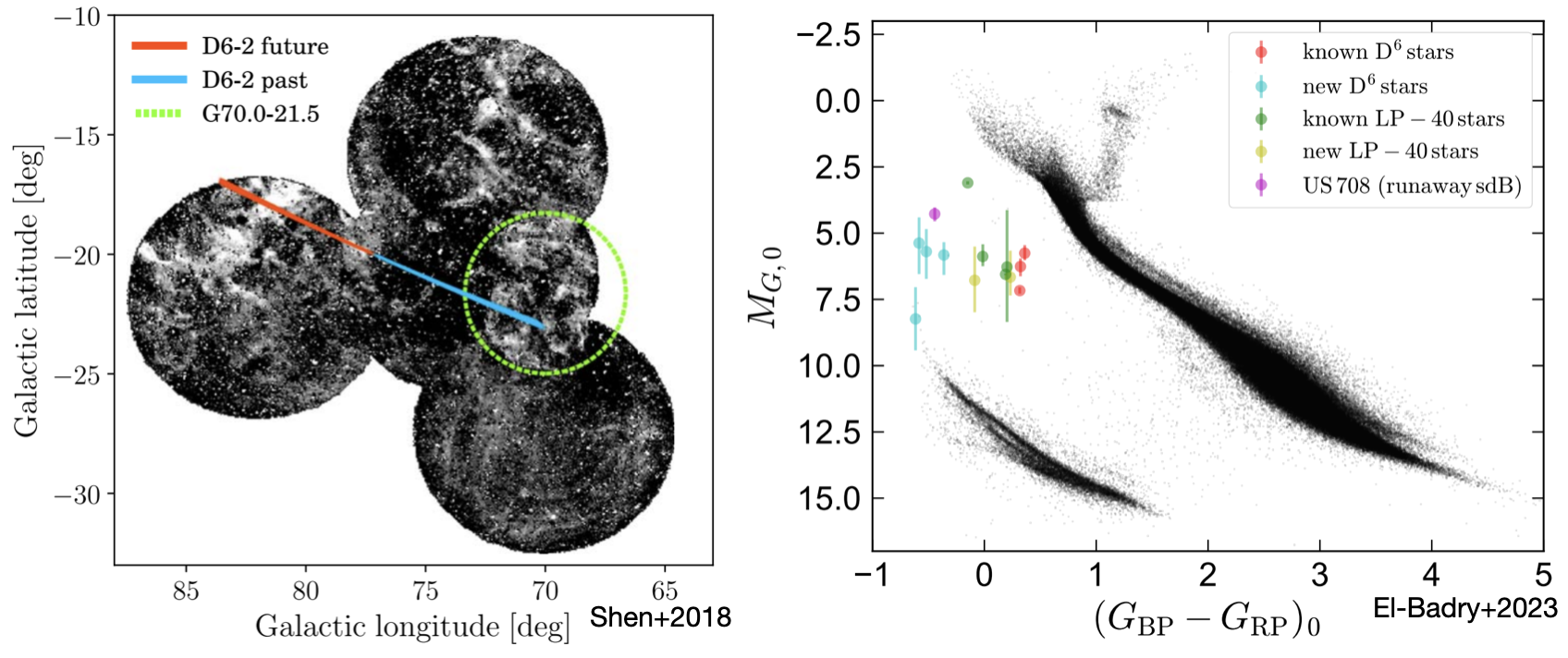}
    \caption{Hypervelocity WDs ejected from thermonuclear supernovae. Left panel, from \citet{Shen2018}, shows the past and future trajectory of the hypervelocity WD D6-2 (one of the red points in the right panel). The trajectory passes through the type Ia supernova remnant G70.0-21.5, with a flight time of $10^5$ yr. Right panel, from \citet{El-Badry2023fastest}, shows the color-magnitude diagram position of the menagerie of hypervelocity WDs and related objects discovered in the last decade  (most of them with {\it Gaia} data).  All these objects sit between the main sequence and the WD cooling track, perhaps because they were inflated during or shortly before the supernova from which they were ejected. }
    \label{fig:d6s}
\end{figure*}
\subsection{Cataclysmic variables and AM CVn binaries}
\citet{Abril2020} and \citet{Abrahams2022} showed that the absolute magnitude of normal cataclysmic variables (CVs) is strongly correlated with orbital period (see Figure~\ref{fig:CVs}). Because the optical luminosity of most CVs is dominated by the accretion disk rather than the donor, this implies that the mass transfer rates decrease with orbital period in an ordered way. While a similar correlation had been found previously based on distances estimated from near-infrared observations of CV donors \citep{Warner1987, Patterson2011}, {\it Gaia} provides more accurate distances and reaches a broader range of CV subtypes.

\citet{Pala2020} assembled a nearly-complete sample of 42 cataclysmic variables within 150\,pc (bottom panel of Figure~\ref{fig:CVs}), from which they measured a CV space density of $\rho = 4.8_{-0.8}^{+0.6}\times 10^{-6}\,\rm pc\,^{-3}$. Volume complete samples are invaluable for the study of heterogenous objects like CVs, since they make it possible to compare the relative frequencies of different subclasses with very different selection functions. \citet{Pala2020} found that magnetic CVs (polars and intermediate polars) make up 36\% of the 150\,pc sample -- a much larger fraction than in earlier CV samples that were not volume limited. 

{\it Gaia} showed a significant fraction of CVs previously thought to be in the 150\,pc sample to be at larger distances. A period gap at $\sim$2-3 hours is clearly seen in the 150\,pc sample (Figure~\ref{fig:CVs}, lower right), but a large majority of CVs in the sample are faint systems with periods below the gap, which are underrepresented in flux-limited samples. Nevertheless, period bouncers -- i.e., CVs that have passed through the period minimum, are evolving toward longer periods, and have degenerate brown dwarf donors -- make up only 7\% of the sample, in strong tension with the predictions of evolutionary models.

{\it Gaia} found the distance of the prototypical AM CVn binary, AM CVn, to be less than half of that previously measured with the {\it HST} fine guidance sensor \citep{Ramsay2018}. This implies that the system has a factor of $\sim 5$ lower mass transfer rate than previously believed, in better agreement with evolutionary models.

\subsection{X-ray binaries}
The {\it Gaia} DR3 distance to the black hole high-mass X-ray binary Cyg X-1, $\varpi = 0.468 \pm 0.015$ mas, is perfectly consistent with the recent VLBI measurement of $0.46\pm 0.04$ mas by \citet{Miller-Jones2021}, which resulted in a revised mass estimate of $21\,M_{\odot}$ for the black hole -- the highest reasonably robust measurement for a BH X-ray binary. {\it Gaia} proper motions and parallaxes also constrained the peculiar velocities of dozens of low-mass X-ray binaries containing black holes and neutron stars \citep{Zhao2023, ODoherty2023}, revealing hints of an anti-correlation between natal kick velocity and compact object mass. Proper motion measurements from {\it Gaia} showed that the low-mass X-ray binary V404 Cyg has a bound wide binary companion at 3500+ au, implying that it formed without a significant kick \citep{Burdge2024}.

\subsection{Contact binaries}
\citet{Hwang2020_cb} used {\it Gaia} astrometry for contact binaries to infer their ages, making use of the fact that the velocity dispersion of stars in the Galactic disk increases with age. They found that contact binaries are relatively old and likely evolved slowly from few-day initial periods by magnetic braking. Linking them to triples (if they have tertiary companions, as e.g. \citealt{Tokovinin2006} find most close binaries do) implies that original inner periods in those triples were of a few days.

\subsection{Hypervelocity white dwarfs}
By using {\it Gaia} astrometry to search for stars with high proper motions and large distances, \citet{Shen2018} identified a population of hypervelocity stars launched from thermonuclear supernovae in close WD binaries. The stars observed today are single, but are believed to be runaways from close WD+WD binaries in which one component explodes and the other flees the scene at a velocity significantly exceeding the Galactic escape velocity. The flight path of the youngest such star identified by \citet{Shen2018} can be traced back to a type Ia supernova remnant (Figure~\ref{fig:d6s}, left panel), in strong support of this formation model.

More hypervelocity white dwarfs were identified by \citet{Raddi2019} and \citet{El-Badry2023fastest}, with velocities ranging from $\sim 500$ to $\sim 2500\,\rm km\,s^{-1}$. \citet{Raddi2019} propose that the slower of these objects (``LP 40-365 stars'') are partially burned runaway accretors from single-degenerate thermonuclear supernovae, while the higher-velocity objects are runaway donors from double-degenerate binaries. All the stars discovered so far sit between the WD cooling track and the main sequence in the CMD (Figure~\ref{fig:d6s}, right panel), though \citet{Gansicke2020} found a somewhat lower-velocity WD with similar abundances that is on the WD cooling track. 

It is, perhaps, not surprising that the objects discovered so far are overluminous: they can be detected at much larger distances than similar objects on the WD cooling track, and their velocities are high enough that they will likely leave the Milky Way before they cool and contract to dimensions of normal WDs. The thermal evolution and expected lifetimes of these objects are still uncertain, however, and models cannot yet simultaneously reproduce their velocities, ages, and luminosities \citep[e.g.][]{Zhang2019, Bauer2019, Bauer2021}. As a result, the birth rate of hypervelocity WDs is quite uncertain. Discovery of these objects has added further evidence that at least a fraction of type Ia supernovae come from double white dwarf binaries, but it remains unclear how large this fraction is \citep[][]{Igoshev2023, El-Badry2023fastest, Braudo2024}.

\section{DR4, DR5, and the road ahead}
The next major data release, {\it Gaia} DR4, is expected in early 2026. DR4 will be transformative for binary star research -- perhaps even more so than DR2 and DR3 -- because it will include epoch-level astrometry for all sources. This will allow self-consistent fitting of {\it Gaia} data and ground-based follow-up and make it possible to fit a variety of different models to the astrometry and RV data. Perhaps most importantly, it will allow the community to explore the regime of lower-SNR astrometric data and experiment with different methods for filtering out spurious solutions, potentially yielding a large sample of BH and neutron star binaries, or other interesting and rare objects. 

Epoch-level astrometric data will, however, represent a significant increase in complexity compared to the clean and mostly well-vetted solutions published so far. 
Epoch-level data in a similar format that will be used in DR4 already exists for a small sample of a few hundred binaries with orbital solutions published by the {\it Hipparcos} mission and can serve as a valuable testing ground for analysis of epoch-level astrometric data while the community waits for DR4. 

There is still significant work to be done in modeling the selection function of various classes of binaries discovered by {\it Gaia}. I proposed in Section~\ref{sec:ast} that astrometric studies of binaries are superior for population modeling because the astrometric signal is easy to forward model. This is true in principle, but it does not yet allow for straightforward modeling of the DR3 binary sample, because a variety of cuts that are challenging to model were applied. 

More than a decade after launch, {\it Gaia} is still observing. Less than a third of the observations {\it Gaia} has already collected have been published so far. The final data release, DR5, is anticipated in 2030 or 2031 and will be based on nearly 11 years of observations -- from mid 2014 until propellant runs out in 2025. These data will foreseeable yield good constraints on the orbits of binaries with periods reaching up to $\sim 20$ years, and useful acceleration constraints at even longer periods \citep[e.g.][]{Andrews2023}. {\it Gaia's} performance at close angular separations -- and thus, its resolution of wide binaries -- is also expected to improve significantly in future data releases \citep{Harrison2023}. The mission's legacy of accurate, all-sky astrometry is unlikely to be superseded for several decades. 

\section*{Acknowledgements}
I thank the reviewers for constructive reports and Andrei Tokovinin and Vasily Belokurov for comments on an early draft.
This research was supported by NSF grant AST-2307232. This work has made use of data from the European Space Agency (ESA) mission {\it Gaia} (\url{https://www.cosmos.esa.int/gaia}), processed by the {\it Gaia} Data Processing and Analysis Consortium (DPAC, \url{https://www.cosmos.esa.int/web/gaia/dpac/consortium}). Funding for the DPAC has been provided by national institutions, in particular the institutions participating in the {\it Gaia} Multilateral Agreement.
%\appendix

%\section{Appendix title 1}

%\section{Appendix title 2}

\bibliographystyle{elsarticle-harv} 
%\bibliography{example}



%% For example:

%% \bibitem[Aladro et al.(2015)]{Aladro15} Aladro, R., Martín, S., Riquelme, D., et al. 2015, \aas, 579, A101

%%\end{thebibliography}

\end{document}